\documentclass[fleqn,usenatbib]{mnras}
\usepackage{newtxtext,newtxmath}
\usepackage[T1]{fontenc}

\DeclareRobustCommand{\VAN}[3]{#2}
\let\VANthebibliography\thebibliography
\def\thebibliography{\DeclareRobustCommand{\VAN}[3]{##3}\VANthebibliography}

\usepackage{graphicx}	
\usepackage{amsmath}	
\usepackage{xspace} 
\usepackage{lineno}
\usepackage{comment}

\newcommand{\be}{\begin{equation}}
\newcommand{\ee}{\end{equation}}

\newcommand{\bi}{\begin{itemize}}
\newcommand{\ei}{\end{itemize}}
\newcommand{\ben}{\begin{enumerate}}
\newcommand{\een}{\end{enumerate}}
\newcommand{\allsize}{1.8}
\newcommand{\cmpsize}{1.5}
\newcommand{\zenodo}{\url{https://zenodo.org/record/7538664}}

\newcommand{\Fermi}{{\textit{Fermi}}}

\title[Multi-class classification of Fermi-LAT sources]{Multi-class classification of Fermi-LAT sources
with hierarchical class definition}

\author[D. Malyshev and A. Bhat]{
Dmitry V. Malyshev,$^{1}$\thanks{E-mail: dmitry.malyshev@fau.de}
Aakash Bhat$^{2}$
\\
$^{1}$Erlangen Centre for Astroparticle Physics, Nikolaus-Fiebiger-Str. 2, Erlangen 91058, Germany\\
$^{2}$Institute of Physics and Astronomy, University of Potsdam, Karl-Liebknecht-Str. 24/25, 14476, Potsdam, Germany
}

\date{Accepted XXX. Received YYY; in original form ZZZ}

\pubyear{2023}

\begin{document}
\label{firstpage}
\pagerange{\pageref{firstpage}--\pageref{lastpage}}
\maketitle

\begin{abstract}

In the paper we develop multi-class classification of \Fermi-LAT gamma-ray sources using machine learning with hierarchical determination of classes. One of the main challenges in the multi-class classification of the \Fermi-LAT sources is that the size of some of the classes is relatively small, for example with less than 10 associated sources belonging to a class. In the paper we propose an hierarchical structure for the determination of the classes. This enables us to have control over the size of classes and to compare the performance of the classification for different numbers of classes. In particular, the class probabilities in the two-class case can be computed either directly by the two-class classification or by summing probabilities of children classes in multi-class classification. We find that the classifications with few large classes have comparable performance with classifications with many smaller classes. Thus, on the one hand, the few-class classification can be recovered by summing probabilities of classification with more classes while, on the other hand, the classification with many classes gives a more detailed information about the physical nature of the sources. As a result of this work, we construct three probabilistic catalogs, which are available online. This work opens up a possibility to perform population studies of sources including unassociated sources and to narrow down searches for possible counterparts of unassociated sources, such as active galactic nuclei, pulsars, or millisecond pulsars.

\end{abstract}





\section{Introduction}

About one third of sources in the \Fermi~large area telescope
(LAT) catalogs are unassociated \citep{2020ApJS..247...33A, 2022ApJS..260...53A, 2010ApJS..188..405A, 2012ApJS..199...31N, 2015ApJS..218...23A}.
Although follow-up observations enable one to find association counterparts at other frequencies for some sources 
\citep[e.g.,][and references therein]{2020ApJS..247...33A, 2022ApJS..260...53A},
the majority of unassociated sources still do not have plausible associations.
Some of these sources may be detectable in gamma rays only. 
For example, a significant fraction (up to 70\%) of pulsars observed in gamma rays are radio quiet 
\citep[e.g.,][]{2016ApJ...833..271S}.
In general, for the gamma-ray sources without counterparts at other frequencies, 
probabilistic classification using machine learning (ML) is the only method to determine 
the most likely physical classes of the sources.

ML algorithms have been used to probabilistically determine the classes of unassociated source by training the ML methods on the associated sources
\citep{2012ApJ...753...83A, 2016ApJ...820....8S, 2016ApJ...825...69M, 2017A&A...602A..86L, 2020MNRAS.492.5377L, 
2021RAA....21...15Z, 2021MNRAS.507.4061F}.
Although there are 23 classes of sources, excluding unassociated sources and sources with unknown physical class%
\footnote{
\label{foot:classes}
The 23 classes of associated sources in the 4FGL-DR3 catalog,
which have a known class of the associated source are \citep{2022ApJS..260...53A}:
gc -- Galactic center,
psr -- young pulsar,
msp -- millisecond pulsar,
pwn -- pulsar wind nebula,
snr -- supernova remnant,
spp -- supernova remnant and/or pulsar wind nebula (both are present at the location of the gamma-ray source),
glc -- globular cluster,
sfr -- star-forming region,
hmb -- high-mass binary,
lmb -- low-mass binary,
bin -- binary,
nov -- nova,
bll -- BL Lac type of blazar,
fsrq -- FSRQ type of blazar,
rdg -- radio galaxy,
agn -- non-blazar active galaxy,
ssrq -- steep spectrum radio quasar,
css -- compact steep spectrum radio source,
bcu -- blazar candidate of uncertain type,
nlsy1 -- narrow-line Seyfert 1 galaxy,
sey -- Seyfert galaxy,
sbg -- starburst galaxy,
gal -- normal galaxy (or part).
}
in the Fourth \Fermi-LAT data release 3 (4FGL-DR3) catalog
\citep{2022ApJS..260...53A},
most of the analyses have been performed for two- or three-class classifications. 
Typical choices of the two classes are extra-galactic and Galactic sources or active galactic nuclei and pulsars. 
Such two- or three-class classifications cannot take into account the rich variety of the different types of gamma-ray sources.

A multi-class classification into more than three classes can give a more detailed information about the possible nature of the 
unassociated sources, which would be useful to narrow down the searches for possible counterparts of the sources and for population studies of the 
different classes of sources.
One of the main challenges for the multi-class classification of the \Fermi-LAT sources is that there are relatively few associated members 
in some of the classes.
For instance, in the 4FGL-DR3 catalog 
\citep{2022ApJS..260...53A}
there are 5 associated or identified star-forming regions, 11 high-mass binaries, 8 low-mass binaries, 4 novae, 2 steep spectrum radio quasars, 8 narrow-line Seyfert 1 galaxies, etc. 
There are also some unique sources, such as the Galactic center.

Although there has been a study of probabilistic classification of unassociated 4FGL-DR3 sources using all 23 physical classes
\citep{2022MNRAS.515.1807C}, it is unclear whether inclusion of classes with very few sources, e.g., less than 10, reduces the stability of classification compared to a classification with fewer classes and it might be beneficial to add the classes with few members to some larger classes.
The main goal of this work is to 
develop a framework for multi-class classification of \Fermi-LAT sources,
which would give a meaningful multi-class classification of sources and provide a comparison of performance of classification
with different numbers of classes.
In particular, we focus on the following question: what is the optimal definition of classes, i.e., which physical classes should be combined in the classification and which ones can enter in the classification as a separate class.

With this question in mind, the main goals of this paper are:
\ben
\item Develop a procedure for class determination;
\item Evaluate performance and check consistency of multi-class classification including a comparison of classifications with different numbers of classes.
\een

The paper is organized as follows. 
In Section \ref{sec:method} we present the details of the data selection and develop an algorithm for the determination of classes
based on Gaussian mixture model (GMM).
In Section \ref{sec:classif} we perform the multi-class classification, evaluate the performance, and compare classifications with different numbers of classes.
In Section \ref{sec:pcat}, we construct the probabilistic catalogs.
Section \ref{sec:concl} contains conclusions and discussion of results.
In Appendix \ref{app:nmin15} we perform classification with groups which have smaller minimal size compared to the classification in Section \ref{sec:classif}, 
in Appendix \ref{app:RFclasses} we determine groups using random forest (RF) rather than the GMM algorithm,
while in Appendix \ref{app:perf_NN} we compare the classification performance for neural networks (NN) versus the RF classification used in Section \ref{sec:classif}.

\section{Data selection and definition of classes}
\label{sec:method}

\subsection{Data selection}
\label{sec:features}

In this work, we use the following 10 features from the 4FGL-DR3 catalog \citep{2022ApJS..260...53A}:
sin(GLAT), 
cos(GLON), 
sin(GLON), 
$\log_{10}$(Energy\_Flux100), 
$\log_{10}$(Unc\_Energy\_Flux100), 
$\log_{10}$(Signif\_Avg), 
LP\_beta,
LP\_SigCurv,
$\log_{10}$(Variability\_Index),
and the index of the log parabola spectrum at 1 GeV%
\footnote{For the definition of the features see \citet{2022ApJS..260...53A}.}.
The last feature is calculated as the derivative at 1 GeV:
\be
{\rm LP\_index}(E) = -\frac{d \log F(E)}{d \log E},
\ee
where $F(E)$ is the log parabola spectrum of the source reported in the 4FGL-DR3 catalog.
Some features are transformed in order to have a comparable range of values. We also use cos(GLON) and sin(GLON) in order to avoid a discontinuity at GLON = $0^\circ$.
The source 4FGL J0534.5+2201i (identified with a pulsar wind nebula) has missing Unc\_Energy\_Flux100, Signif\_Avg, and Variability\_Index values
in the 4FGL-DR3 catalog. This source is excluded from the analysis in this paper.

These features characterize the main properties of gamma-ray sources, such as the position on the sky,
the shape of the energy spectrum, overall flux, and variability as a function of time.
We have also checked in an earlier analysis that these features have relatively small correlation among themselves, i.e., they are not redundant
\citep{2022A&A...660A..87B}.
In principle, not all 10 features may be important for the classification. 
However, it is not possible to do such selection of features a priori.
For a given set of classes and a classification algorithm, one can determine an optimal number of features by removing the least important feature one at a time \citep{2020MNRAS.492.5377L}.
Such a study goes beyond the scope of this work, where we focus on the effects of the determination of classes and use the same input features for a consistent comparison of the different cases.

\subsection{Definition of classes}
\label{sec:class_def}

The 4FGL-DR3 catalog has 23 classes of sources \citep{2022ApJS..260...53A}.
We combine identified and associated sources and use the lower case class labels for the corresponding classes 
(see footnote~\ref{foot:classes}).
There are also associated sources with an ``unknown'' class (unk): 
these are sources at latitudes $|b| < 10^\circ$ with an association at other wavelengths based on likelihood ratio method only,
i.e., no significant association with the Bayesian method is found \citep{2022ApJS..260...53A}.
As a result the association class is considered to be uncertain.
We add the sources with an unknown class to the unassociated sources.

In this section we determine the division of physical classes into groups, which have a good separation in the feature space.
The determination of the groups proceeds iteratively: we first divide the classes into two groups, 
then each of the groups is sub-divided into two groups and so on until one of the terminations condition is met.
For the division we use the GMM.
The procedure is as follows:
\ben
\item
\label{it:step1}
We model the distribution of all sources in the feature space by the GMM with two kernels labeled as ``0'' and ``1''. 
For a source $i$ we denote by $p^i_k$ the probability that the source belongs to the distribution given by kernel $k = 0,\,1$.
Since there are only two kernels, $p^i_0 = 1 - p^i_1$.
\item
\label{it:step2}
For all sources in a physical class ``m'' we compute an average probability that the sources belong to kernel $k = 0,\,1$: 
$\bar{p}^m_k = \sum_{i\in m} p^i_k / N_m$, where $N_m$ is the number of sources
in class $m$. If the average group-1 probability for class $m$ is larger than the average $p_1$ probability 
for all associated sources $\bar{p}_1 = \sum_i p^i_1 / N$, i.e., $\bar{p}^m_1 > \bar{p}_1$, 
then class $m$ is added to group 1, otherwise it is added to group 0.
We compare the group-1 probability to the average probability over all sources in order to make the subgroups as balanced as possible.
\item 
\label{it:term}
Steps \ref{it:step1} and \ref{it:step2} are repeated iteratively, until one of the termination conditions is met: (1) the group contains only one physical class, (2) a further sub-division of a group would lead to a group with less than a certain minimum number of sources, (3) a maximum number of sub-division is reached.
In the baseline model, we require that the number of sources in a subgroup $n_{\rm min}$ is not less than 100 and we do not put constraints on the number of sub-divisions.
\een

In Appendix \ref{app:nmin15}, we present the determination of the groups with a smaller minimal number of sources, $n_{\rm min} > 15$, and with the maximum of 4 subdivisions.
We also use a supervised learning approach with the RF algorithm for the definition of the groups in Appendix \ref{app:RFclasses} instead of the unsupervised GMM approach in this section.

After the first division, the classes are grouped as follows: \\
Group 0: sbg, rdg, bll, css, bcu, ssrq (3009 sources); \\
Group 1: fsrq, sey, nlsy1, agn, glc, nov, spp, msp, lmb, hmb, psr, sfr, snr, gc, pwn, gal, bin (1358 sources).

\begin{figure}
\includegraphics[width=\columnwidth]{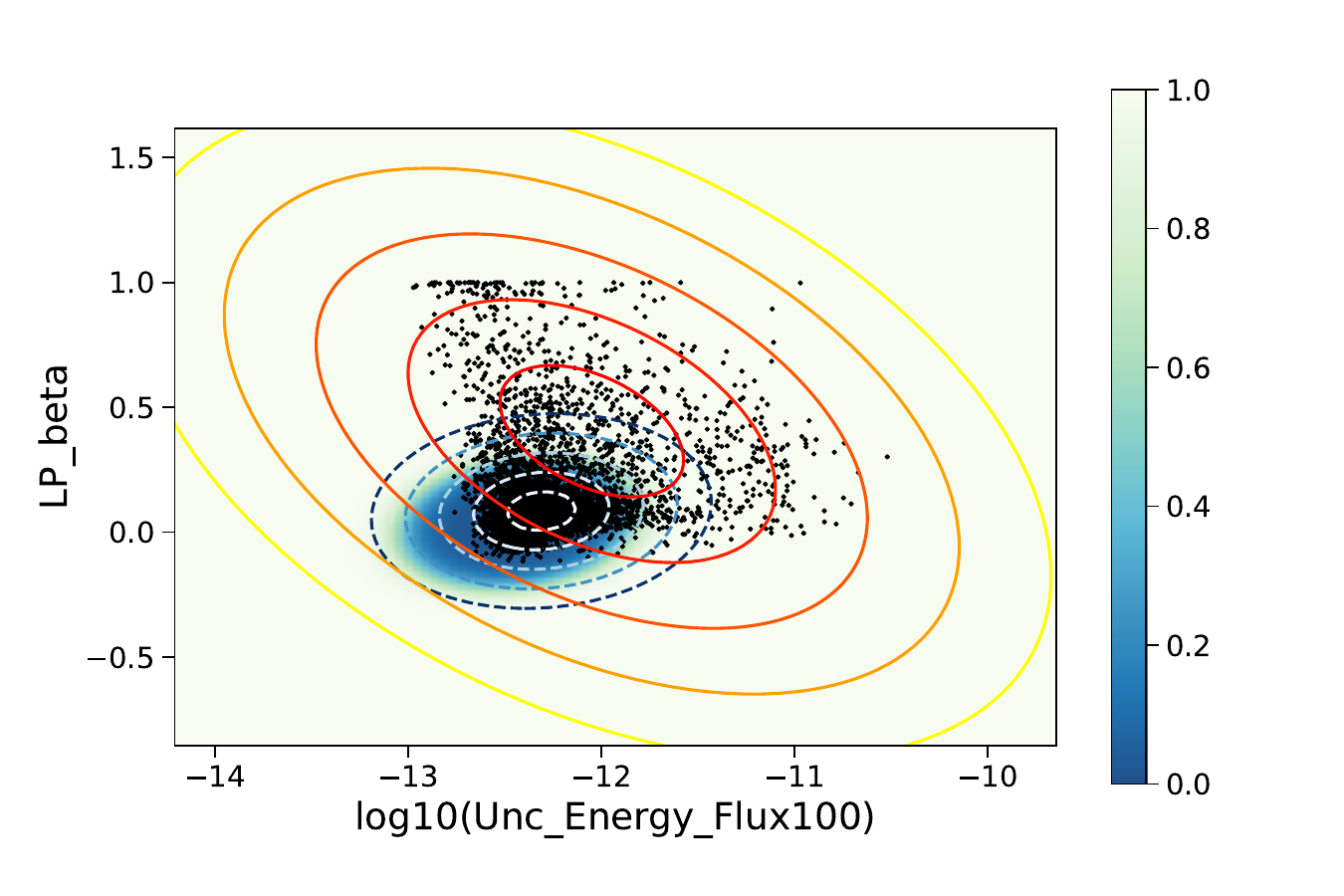}
\includegraphics[width=\columnwidth]{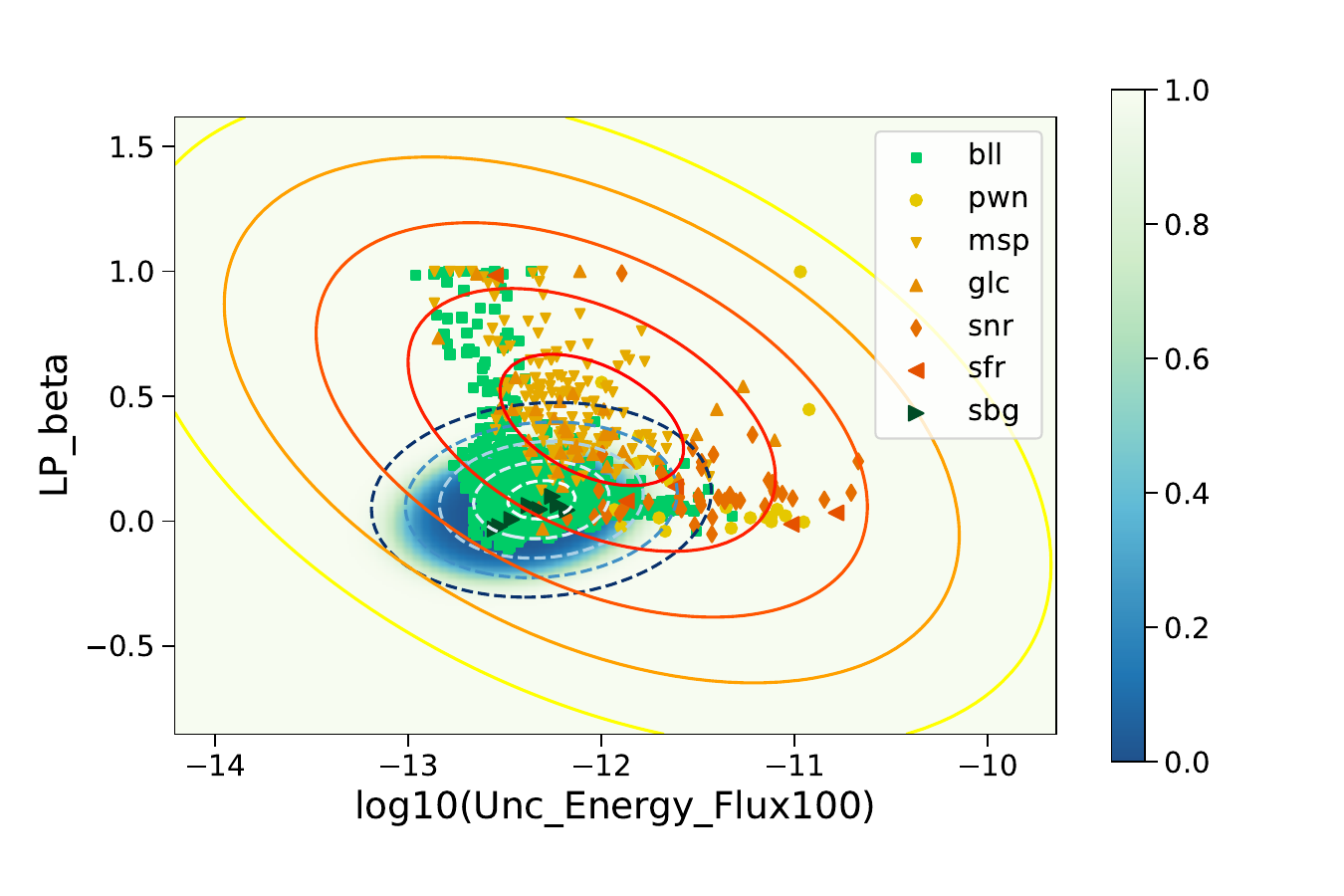}
\caption{Separation of 4FGL-DR3 associated PS into two groups using GMM with two Gaussian kernels.
Top panel: all associated PS and the $1, 2, \ldots, 5$ sigma contours of the two Gaussian kernels (dashed lines - kernel 0, solid lines - kernel 1).
Color scale: probability for a source to belong to kernel 1 distribution.
Bottom panel: examples of distributions of physical classes attributed to kernel 0 {(bll, sbg)} or kernel 1 {(pwn, msp, glc, snr, and sfr)}.
A physical class is attributed to kernel 1 if the average kernel 1 probability of the class members are larger than the average kernel 1 
probability for all associated sources (see text for more details). The distributions in this plot are shown for two features.
The actual distribution of the physical classes into groups and classification of sources is performed in the 10-dimensional feature space
described in Section \ref{sec:features}.
}
\label{fig:node0_split_black}
\end{figure}

We show in Fig.~\ref{fig:node0_split_black} an example of the GMM model in two features, the log-parabola curvature parameter
(LP\_beta) and the logarithm of the energy uncertainty $\log_{10}$(Unc\_Energy\_Flux100).
These two features are the most important features for the two-group separation by the GMM in the 10-dimensional feature space,
where we estimate the importance by the ratio of the distance between the means of the Gaussians
divided by the square root of the sum of sigmas squared:
\be
S(f) = \frac{abs(\mu(f)_1 - \mu(f)_0)}{\sqrt{\sigma^2(f)_1 + \sigma^2(f)_0}}.
\ee
For the illustration in Fig.~\ref{fig:node0_split_black}, we construct a 2-dimensional GMM, i.e.,
the model is different from the 10-dimensional GMM used for the separation of classes.
In this particular case, however, the 2D and the 10D GMM models give the same separation of classes into 2 groups.

Dashed and solid contours in Fig.~\ref{fig:node0_split_black} show the 1, 2, 3, 4, and 5 sigma levels of the Gaussian kernels 0 and 1 respectively.
Top panel shows the distribution of all 4FGL-DR3 sources.
On the bottom panel we show several classes with the corresponding separation into groups:
green points show a class that belongs to group 0 {(bll, sbg)}, while red or orange points show the classes belonging to group 1
{(pwn, msp, glc, snr, and sfr)}.
The color scale shows the probability that a source belongs to group 1 according to the GMM.

{
The separation is performed based on similarities of the gamma-ray spectra and other properties of sources, such as variability and position on the sky.
As a result, sources with similar gamma-ray spectra and other properties will be attributed to the same group of classes even if the 
gamma-ray production mechanisms are different.
For example, the starburst galaxies (sbg) are added to Group 0, which mostly contains blazars: BL Lacs (bll) and blazars of unknown type (bcu),
whereas gamma-ray production mechanism is similar to some of the classes in Group 1, such as supernova remnants (snr) and star-forming regions (sfr).
Nevertheless, the gamma-ray properties of sbg sources make them look similar to BL Lacs (featureless power-law spectra, absence of variability, 
isotropic distribution on the sky). 
In particular, positions at high latitudes result in low flux uncertainty for sbg sources similar to bll sources, 
which distinguishes them from from snr and sfr sources 
(see Fig.~\ref{fig:node0_split_black} lower panel).
The attribution of physical classes to groups also depends on the method used to separate the classes into groups.
For example, in Appendix \ref{app:RFclasses} we use the RF algorithm for the definition of the groups. 
In this case, snr and gal (normal galaxy or part) classes are in the same group with sbg and bll classes.
}

We note that the sub-division in groups 0 and 1 has a structure of a tree where 0 corresponds to a node on the left and 1 corresponds to a node on the right of a parent node.
Thus, the first subdivision can be written as a tree with three nodes: 0 - root node that contains all classes, and two children nodes 00 (corresponding to group 0) and 01 (corresponding to group 1).
In other words, after the first division the tree has one internal node (the root 0) and two external nodes or leafs (00 and 01).

After the second division, the groups denoted by the corresponding external nodes are:\\
000: sbg, rdg, bll, css (1514 sources); \\
001: bcu, ssrq (1495 sources); \\
010: fsrq, sey, nlsy1, agn (813 sources); \\
011: glc, nov, spp, msp, lmb, hmb, psr, sfr, snr, gc, pwn, gal, bin (545 sources).

\begin{figure}
\includegraphics[width=\columnwidth]{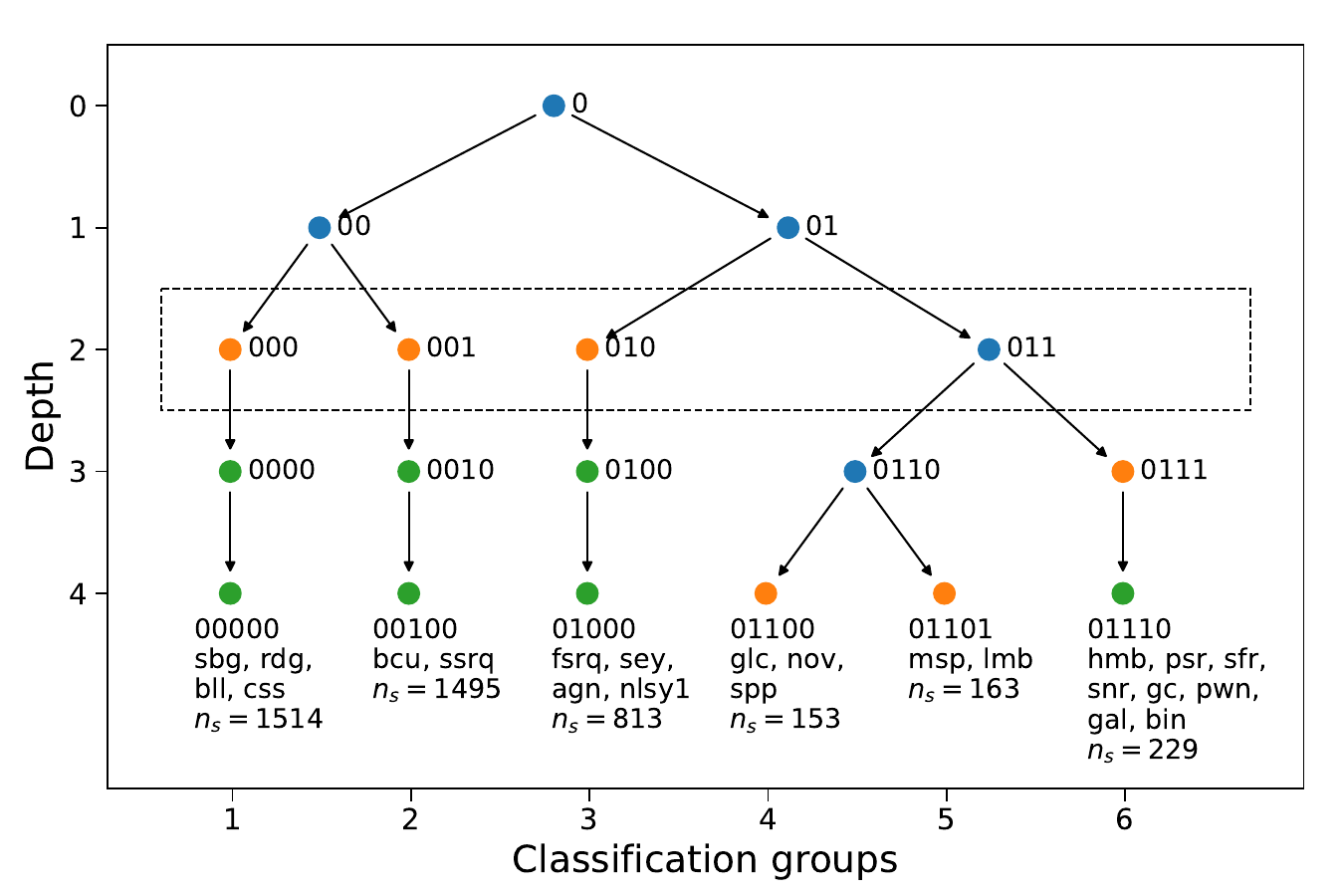}
\caption{Hierarchical definition of groups of physical classes of the Fermi-LAT 4FGL-DR3 catalog
determined with the GMM.
The physical classes are written below the nodes at Depth = 4. 
$n_s$ is the total number of associated sources in all physical classes at the node.
The definition of class names can be found in footnote \ref{foot:classes}.
Blue nodes -- groups which have further splits. 
Orange nodes -- final groups (no further splits due to one of the termination conditions).
Green nodes -- copy of orange nodes to the next depth for visualization.}
\label{fig:tree}
\end{figure}

We continue the subdivision process until it stops in all nodes according to the termination conditions in \ref{it:term}, which in our case corresponds to 4 steps.
The results of the subdivision are represented as a tree in Fig.~\ref{fig:tree}.
The y-axis shows the depth level, which is the number of divisions.
The name of a node, consisting of 0's and 1's, shows the path to the node from the root: 0 - left, 1 - right.
The root of the tree contains all 23 physical classes of the 4FGL-DR3 catalog. 
Blue nodes show groups which have further splits. Orange nodes are the final groups, which have no further splits due to one of the termination conditions. 
If a group is obtained by fewer than four splitting steps (the maximal number of splits in this case), 
then in order to have all physical classes represented at each depth level,
we copy the group to the next level and add 0 at the end of the node name. 
The corresponding copies of nodes are represented by green circles. 
The physical classes in the final groups are listed at the bottom of the tree (at depth four).
For example, the group at the bottom left node ``00000'' has four physical classes: sbg, rdg, bll, and css, which in total have 1514 sources in the 4FGL-DR3 catalog.
For a node at depth smaller than four the physical classes are determined by adding the physical classes of its children nodes at depth four.
As a result of this procedure, all physical classes are present at each depth, but they are combined in groups of different sizes at the different depths.

\section{Multi-class classification}
\label{sec:classif}

In this section we use the groups of classes defined in Section~\ref{sec:class_def} for 
the classification of \Fermi-LAT sources.
The number of groups of classes is determined by the number of steps in the division process,
i.e., the depth of the tree in Figure~\ref{fig:tree}.
We consider four classification problems given by the groups at depths one to four respectively.
In Figure~\ref{fig:tree} we show the groups used in the classification at depth two by a dashed rectangle.
The physical classes in the groups are obtained by adding the classes in the children nodes at depth four. 
For example, physical classes in node 0110 at depth three come from nodes 01100 and 01101 at depth four
(the classes are: glc, nov, spp, msp, lmb).

For the classification, we use the RF
and NN algorithms%
\footnote{We use the RF and NN algorithms implemented in scikit-learn 1.0.2 \citep{scikit-learn}
as RandomForestClassifier and MLPClassifier respectively.
For the RF algorithm we optimize the Gini index, while for the NN we optimize the log-loss function (cross-entropy).}.
We show the results for the RF algorithm with 50 trees with the maximal depth of 15 in this section.
The results for the NN algorithm are presented in Appendix \ref{app:perf_NN}.
We split the associated sources into training and testing samples with the ratio 70/30\%.
We use the training sample to train the classification algorithm, while the testing samples are used for the calculation of the performance.

We compare performance of classification with different numbers of classes
using receiver operating characteristic (ROC) curves, precision and recall, and reliability diagrams. 
One of the questions that is addressed with the hierarchical definition of groups of classes is whether the subdivision of groups improves the classification, 
e.g., due to additional information that is contained in the sub-groups, 
or makes the classification worse, e.g., due to confusion of ML classification algorithms in the presence of many classes.

\subsection{ROC curves}

\begin{figure}
\includegraphics[width=\columnwidth]{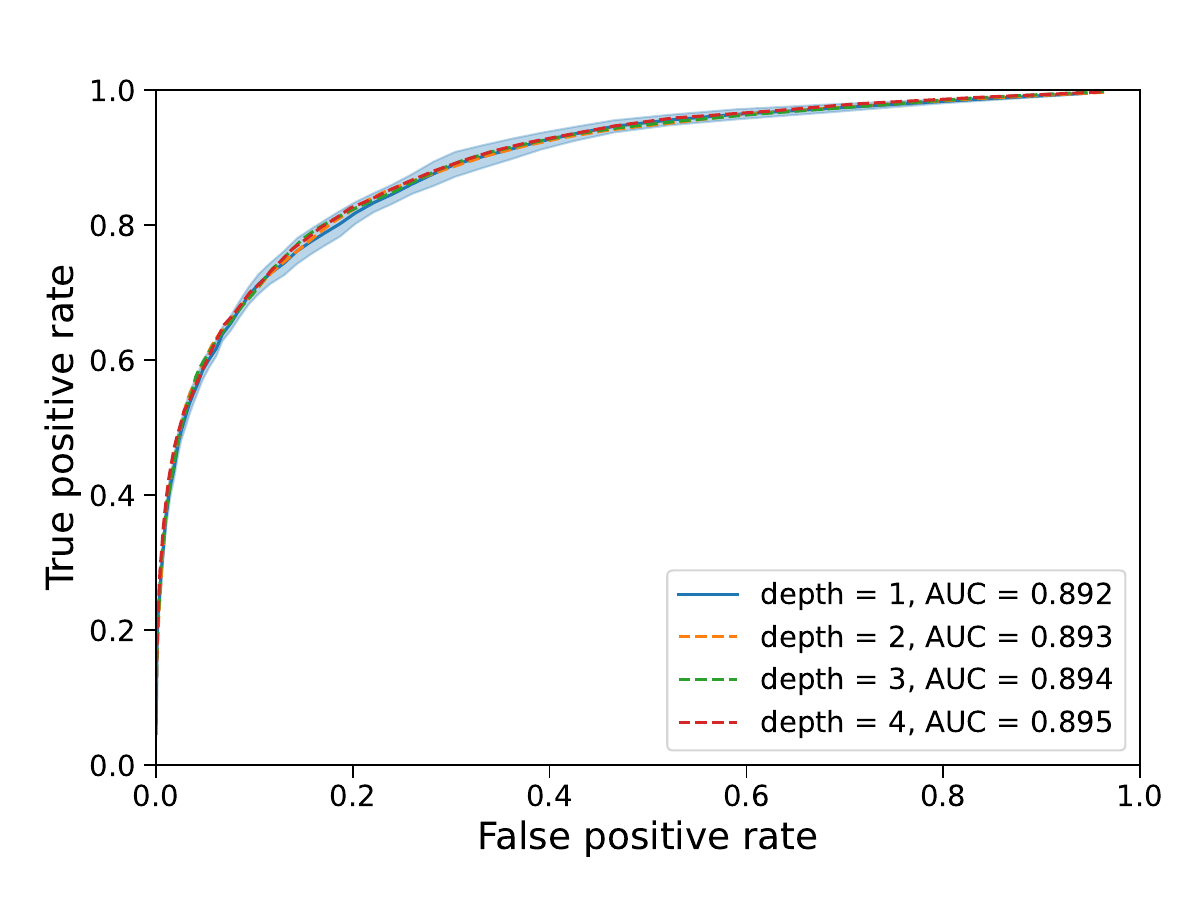}
\caption{ROC curves for the 01-class in Figure \ref{fig:tree}.
The probabilities are calculated either directly in the two-class classification with 00 and 01 classes (solid blue line)
or by summing the probabilities of the children nodes of the 01-node in the multi-class classification
at depth 2, 3 and 4 shown by dashed orange, green and red lines respectively. The blue shaded region shows the statistical uncertainty
for the direct two-class classification obtained by 10 random splits into training and testing samples.
}
\label{fig:ROC}
\end{figure}

As a first test we compute the ROC curves.
We start by calculating the ROC for the two-class classification problem.
The results for the 01 class are shown in Fig.~\ref{fig:ROC}.
The different lines correspond to different ways to compute the probability of the 01 class.
At depth 1, the probability is directly calculated in the two-class classification including 00 and 01 classes (see Fig.~\ref{fig:tree}).
The blue band shows an estimate of the statistical uncertainty obtained by 10 splits into training and testing sets in the two-class classification.
At depth 2, we perform the 4-class classification with classes 000, 001, 010, and 011 
(shown by dashed rectangle in Fig.~\ref{fig:tree}) and then sum the probabilities of the children classes of 01 node (e.g., 010 and 011 classes) 
in order to determine the 01-class probability.
Analogously, by summing the probabilities of the children nodes of the 01-class at depths three and four in Fig.~\ref{fig:tree}, 
we compute the 01-class probabilities corresponding to these depths.
We find that the ROC curves in this case are consistent for the classifications with the different numbers of classes, 
e.g., the classifications give very similar area under curve (AUC) shown in the labels in Fig.~\ref{fig:ROC}.
It means that including more classes does not reduce the two-class classification performance.
Similar conclusions about the classification performance have been obtained 
for a comparison of two- and three-class cases \citep{2022arXiv220910236M}.

\begin{figure*}
\includegraphics[width=\allsize\columnwidth]{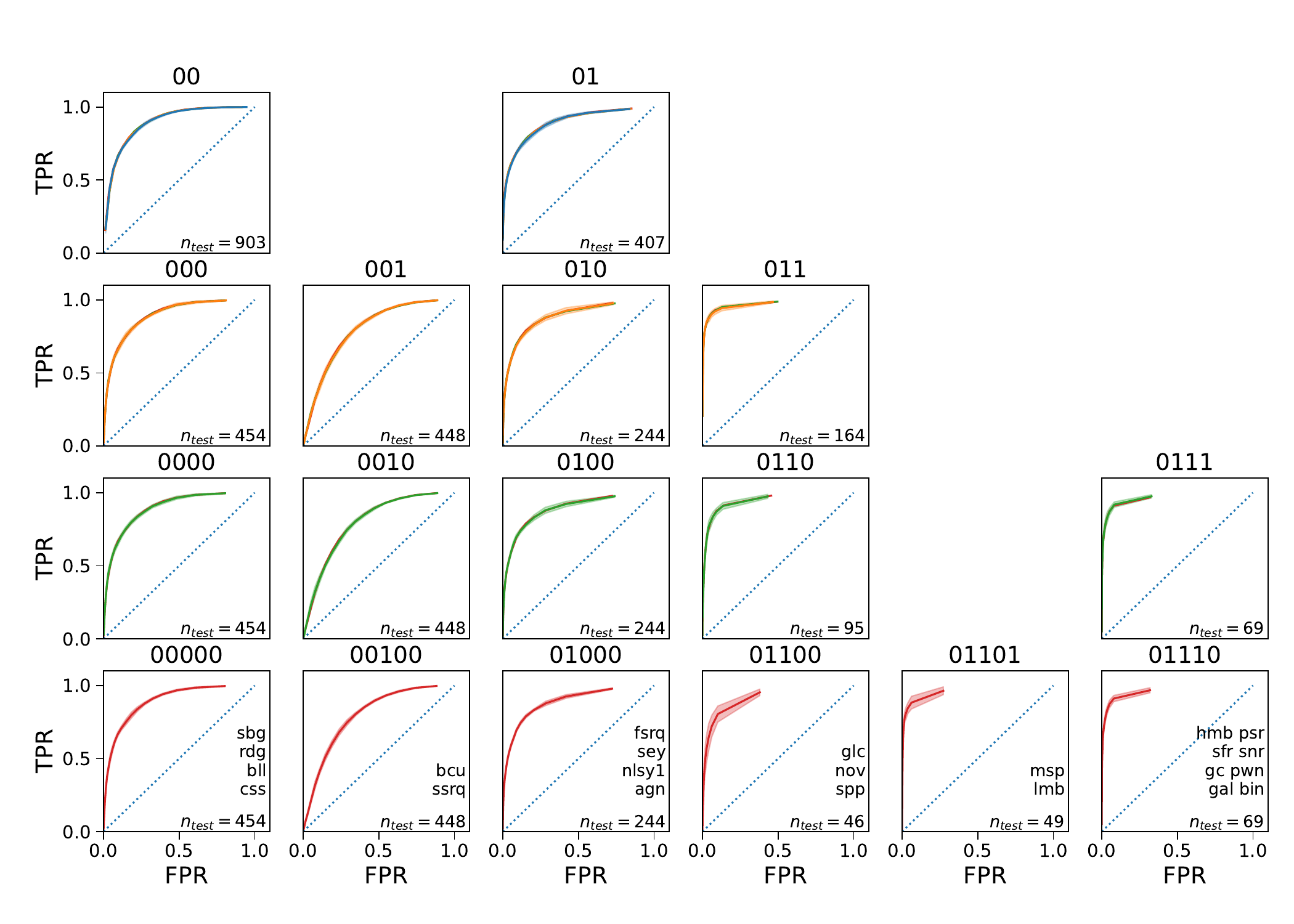}
\caption{ROC curves for all groups in Figure \ref{fig:tree}. Similarly to Figure \ref{fig:ROC}, 
we calculate the ROC curves either directly or by summing the probabilities of the children nodes using the multi-class classification at larger depths.
Red lines in the bottom panels show the results of the direct six-class classification.
We also use the red lines to show the results for summation of class probabilities in the six class classification for
the calculation of ROC curves at lines one (two-class classification), two (four-class classification), and three (five-class classification).
Similarly green (orange) lines show the results of direct five-class (four-class) classification in row three (two) and the results of class probabilities summation
at higher levels.
Blue lines at the first row show the results of the direct two-class classification.
Shaded regions show the statistical uncertainty in the direct two, four, five, or six-class classifications obtained by 10 random splits into training and testing samples.
All lines are very close to each other, which means that direct classification gives similar results as classification with more groups and summation of probabilities of children nodes.
$n_{\rm test}$ is the number of test sources used for the calculation of the ROC curves.
The physical classes of the groups are shown at the bottom raw.
}
\label{fig:ROC_all}
\end{figure*}

In Fig.~\ref{fig:ROC_all} we show the ROC curves for all groups represented by nodes in Fig.~\ref{fig:tree}.
The position of the panel in Fig.~\ref{fig:ROC_all} reflects the position of the group in the tree in Fig.~\ref{fig:tree}.
For instance, in the first row of Fig.~\ref{fig:ROC_all} there are two groups corresponding to the nodes 00 and 01,
in the second row there are four groups 000, 001, 010, and 011, 
where the children nodes of the 00 node (000 and 001) are either below or to the right of this node.
Analogously, the children nodes of the 01 node (010 and 011) are below and to the right of this node.

For the ROC curves at the last row in Fig.~\ref{fig:ROC_all}, we use six-class classification with the corresponding groups of classes.
The red band in these panels shows an estimate of the statistical uncertainty obtained by 10 splits into training and testing samples.
We use the one-vs-all calculation of true and false positive rates in the multi-class classification.
In the third row, the ROC curves are calculated in two different ways: either by directly calculating the class probabilities in the five-class classification
with the corresponding groups of classes (green curves) or by summing the probabilities of the children nodes in the last row (red curves).
Since only one group is divided between row three and four (0110 is divided into 01100 and 01101),
we sum probabilities for groups 01100 and 01101 for 0110 and use the 6-class probabilities for the other groups for the red curves in row three.
The red curves follow very closely the green curves and are hardly visible, which means that (at least in the ROC curves case)
the six-class and the five-class classifications are equivalent from the point of view of the five-class probabilities.
The six-class classification, however, provides more detailed information about the 0110 group by dividing it into two subgroups.
The green bands show the estimated statistical probability in the direct
five-class classification obtained by the 10 random splits into training and testing samples.

Analogously, the ROC curves in the first and second rows in Fig.~\ref{fig:ROC_all} are calculated either directly by
two- or four-class classifications or by summing the probabilities of the children nodes at lower levels.
For example, in the four-class case, the four-class probabilities obtained by summing the probabilities of the five-class (six-class) classification are shown
by green (red) lines in the second row, while the probabilities calculated directly in the four-class classification are shown by 
orange lines (the orange bands show the statistical uncertainty in the direct four-class classification obtained by 10 random splits into 
training and testing samples).
In the two-class case, the direct two-class probabilities are shown by blue lines (the corresponding statistical uncertainty is shown
by the blue bands), while the two-class probabilities obtained by summing the groups at levels 2, 3, and 4 are shown by orange, green, and red lines respectively.

We see that in all cases, direct classification gives results comparable with summing probabilities of classifications with more groups.
Thus, six-class classification has an advantage in comparison with classifications with fewer classes, because the two, four, and five-class probabilities 
can be calculated from the six-class probabilities by summing the probabilities of the corresponding subgroups with similar performance as the direct classification with fewer groups, while the six-class classification gives more detailed information about the classes of the sources.

\subsection{Precision and recall}

\begin{figure}
\includegraphics[width=\columnwidth]{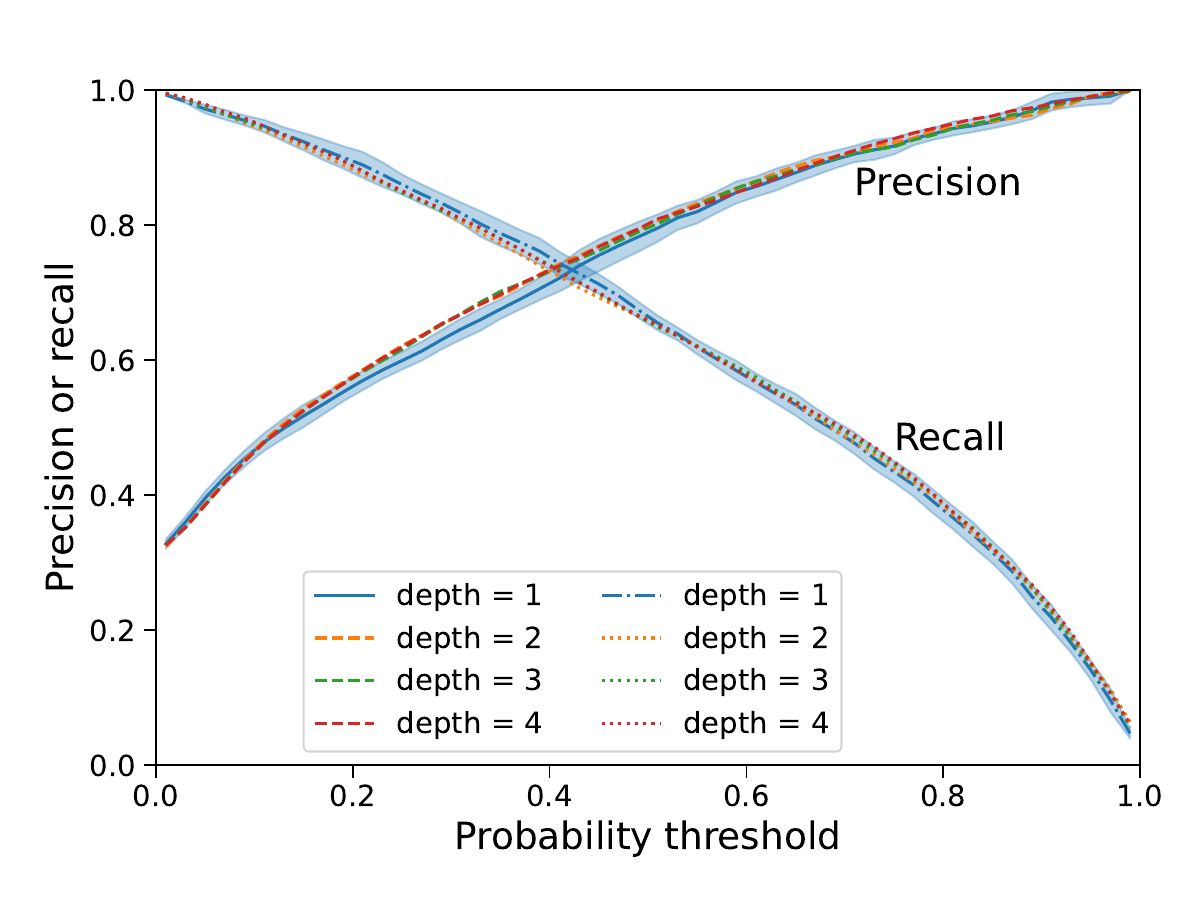}
\caption{Precision and recall for the 01-class in Figure \ref{fig:tree} as a function of the 01-class probability threshold.
The lines are calculated either directly using the two-class classification 
(solid blue line -- precision, dash-dotted blue line -- recall) or by summing the class probabilities of the children nodes of the 01 node
at depths two, three, and four: dashed (dotted) orange, green, and red lines for precision (recall) respectively.
}
\label{fig:prec_recall}
\end{figure}

In this section we compare the performance of classification with different numbers of groups using precision and recall.
Precision (or purity) is the fraction of true positive class candidates relative to all class candidates,
while recall (or completeness) is the fraction of true positive candidates to all true class members.
Given a probability threshold $p_{\rm thr}$ the positive candidates for class $m$ are determined by the 
condition $p^m_i > p_{\rm thr}$.
The precision and recall for the class corresponding to group 01 as a function of probability threshold $p_{\rm thr}$ for two-class classification
are shown in Fig.~\ref{fig:prec_recall}.
Similarly to the ROC curves in Fig.~\ref{fig:ROC}, the blue solid and dash-dotted lines show respectively the precision and recall for the direct two-class classification.
The blue shaded areas show respectively the statistical uncertainties obtained with 10 random splits into training and testing datasets.
The orange, green, and red dashed (dotted) lines show the precision (recall) for probabilities obtained by summing 
the probabilities of the children nodes in the four, five, and six-class classifications at depths two, three, and four respectively.
Similarly to the ROC curves, the precision and recall curves obtained with the direct two-class classification and by summing the probabilities
of multi-class classifications are very similar.

\begin{figure*}
\includegraphics[width=\allsize\columnwidth]{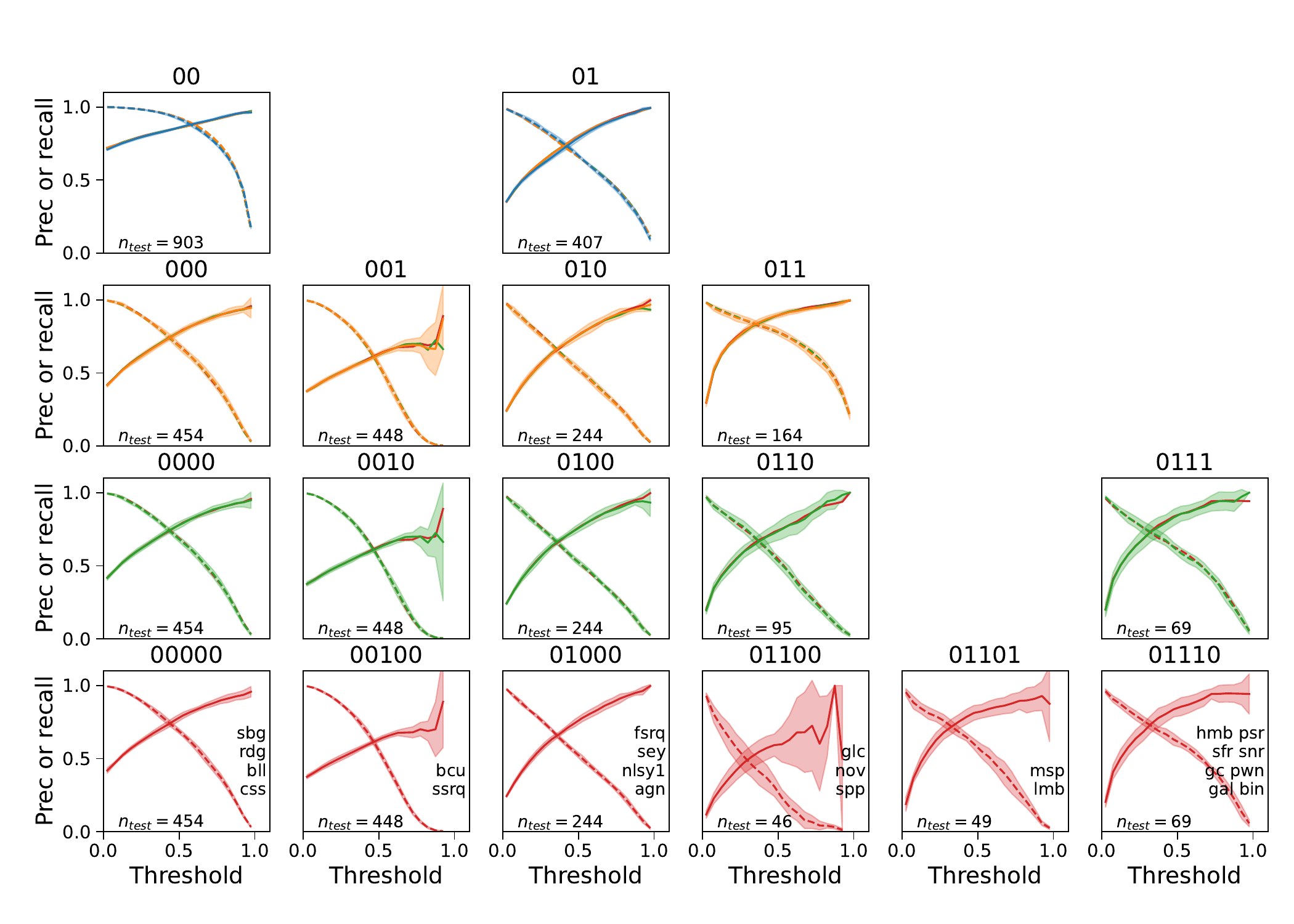}
\caption{Precision and recall for all groups in Figure \ref{fig:tree}. The lines are defined similarly to Figure \ref{fig:ROC_all}.
Solid (dashed) lines show precision (recall).}
\label{fig:prec_recall_all}
\end{figure*}

In Fig.~\ref{fig:prec_recall_all} we present the calculation of precision and recall either directly using two, four, five, and six-class classification shown by blue, orange, green and red lines  in rows one, two, three, and four respectively or by summing the probabilities of the children nodes.
For example, the precision and recall obtained with the six-class classification in the two, four, and five-class cases are shown by red lines in the rows one, two, and three respectively. Analogously, the five-class (four-class) precision and recall are shown by green (orange) lines.
We see that direct classification has similar performance as the classification obtained by summing the probabilities of the children nodes.
In particular, we find that the precision and recall as a function of the probability threshold are similar either with direct two, four, or five-class classifications or by summing the probabilities of the children nodes in the six-class classification.

\subsection{Reliability diagrams}

\begin{figure*}
\includegraphics[width=\allsize\columnwidth]{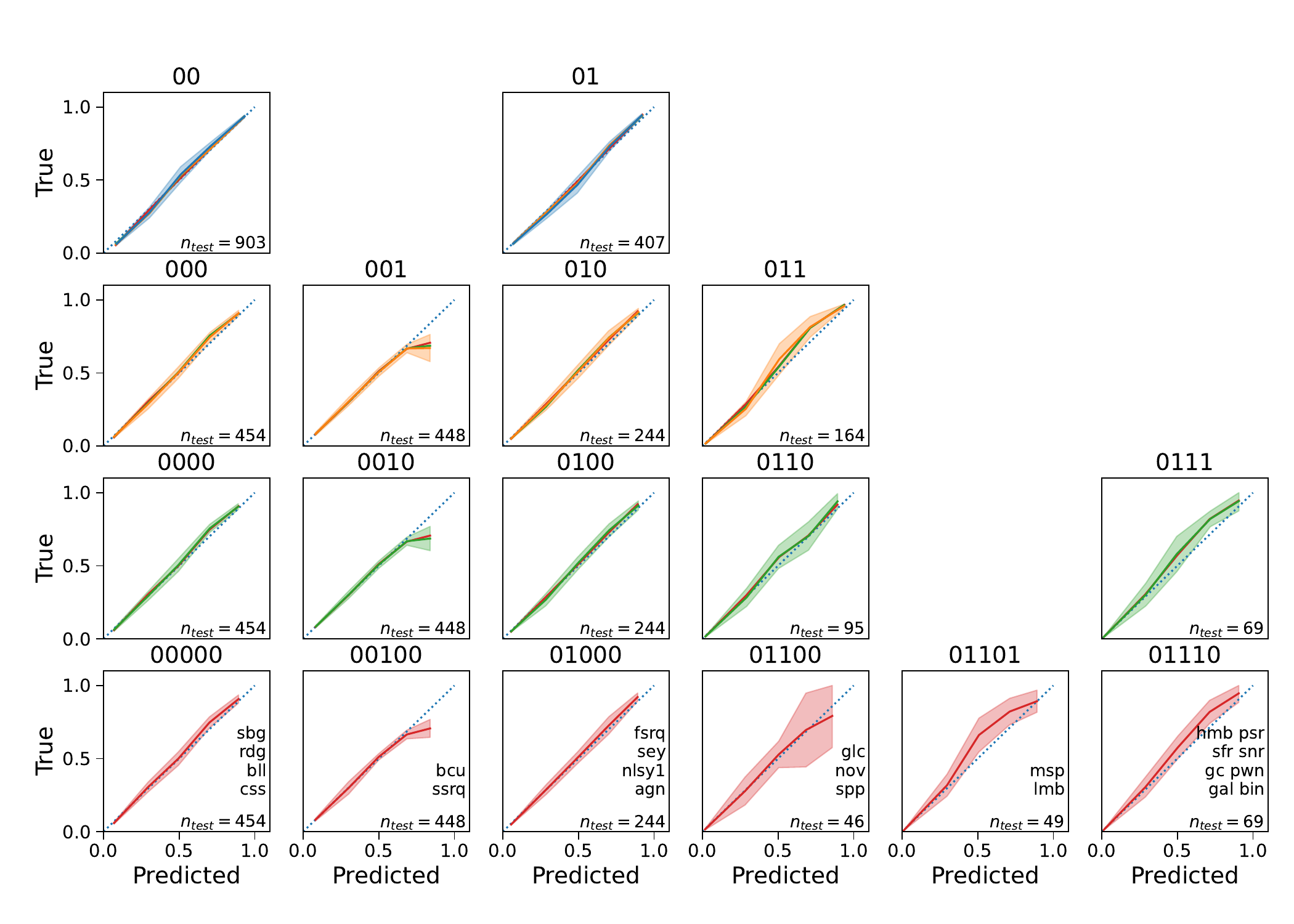}
\caption{Reliability diagrams for all groups in Figure \ref{fig:tree}. The lines are defined similarly to Figure \ref{fig:ROC_all}.
Dotted blue lines show the $y = x$ line (perfect calibration).}
\label{fig:reliability}
\end{figure*}

In this subsection we check that the class probabilities given by the RF algorithm are well calibrated, i.e., 
they can be interpreted as probabilities that the class candidates are true members of the corresponding classes.
In Fig.~\ref{fig:reliability} we show the reliability (or calibration) diagrams for the classifications with different numbers of classes.
A reliability diagram is calculated by taking all class candidates in a range of class probabilities (the probability threshold on the x-axis)
and by calculating the fraction of the true class members among the candidates in this range of class probabilities (the true fraction on the y-axis).
The well calibrated probabilities should be close to the optimal case given by the $y = x$ line (shown by dotted lines on the panels).
Similarly to the ROC curves in Fig.~\ref{fig:ROC}
and to the precision and recall in Fig.~\ref{fig:prec_recall_all}, 
the blue lines in the first row show the reliability for the direct two-class classification.
The orange, green, and red lines show the reliability for probabilities obtained by summing 
the probabilities of the children nodes in the four, five, and six-class classifications at depths two, three, and four respectively.
The reliability diagrams in the second, third and fourth rows show either direct classification into four, five, and six classes (orange, green, and red lines respectively)
or by summing the probabilities of the children nodes.
The shaded areas show the statistical uncertainties obtained with 10 random splits into training and testing datasets for the direct classification
into two, four, five or six classes.
We find that in all cases the probabilities are well calibrated within the statistical uncertainties.

\subsection{Systematics tests}

Overall, we find that the six-class classification probabilities can be used to calculate the classification
probabilities for two, four, or five class classifications by summing the class probabilities of the children nodes without the loss of performance,
while they also give more detailed information about the classes of the sources compared to the classifications with less than six groups.
We consider the multi-class classification using RF algorithms for the six groups at depth four in Fig.~\ref{fig:tree} as a baseline model.

In Appendix \ref{app:nmin15} we compare the baseline model to a classification into nine groups derived with the GMM algorithm with 
the condition $n_{\rm min} > 15$ and the maximal number of subdivisions equal to four.
We find that similar to the six-class classification, the classification with nine groups has similar performance as classification with fewer groups obtained by joining
some of the nine groups.
In case of nine groups, however, some of the groups are so small that the classification algorithm does not find any reliable candidates, 
i.e., the groups consisting of rdg, css and agn, nlsy1, sey classes, where precision and recall values are very small.
We also compare the performance of classification with six and nine classes by comparing the 
ROC, precision, and recall for the groups dominated by one of the large classes,
e.g., bcu, bll, fsrq, msp, psr, and spp (we note that there is only one of these large classes in any of the six groups at depth four in Fig.~\ref{fig:tree}).
We find that the performance for the six large groups is comparable for the six- and nine-class classifications, 
while there is one more group in the nine-class case consisting of snr, pwn, and gc classes, which also has reasonable ROC, precision, and recall.

In Appendix \ref{app:RFclasses} we use RF algorithm for the definition of the sub-division of groups.
The algorithm has the same iterative structure as the group sub-division algorithm described in Section \ref{sec:class_def},
but instead of the GMM with two Gaussian kernels we use RF with two classes.
The two classes (0 and 1) are determined by using two largest physical classes in the group to train the RF algorithm.
Then all physical classes in the group are classified according to the RF classification.
The condition to attribute a physical class to a sub-group is the same as in the GMM case: if the average probability of members in a physical class
is larger than the average group-1 probability for all sources in the parent group, then the physical class is attributed to group 1,
otherwise it is attributed to group 0.
We use the condition for the minimal number in a group $n_{\rm min} > 100$, which gives six final groups, as in the case of the GMM algorithm, but with a different distribution of physical classes among the groups.
The performance with the groups determined by the RF algorithm is similar or better than the performance for the groups defined with the GMM algorithm
apart for one group, which consists of msp and lmb in the GMM case and msp and glc in the RF case (for this group the performance of the GMM-defined group is better).
This is not surprising, since we also use the RF algorithm for the multi-class classification.
Overall, the differences in the performance are relatively small.

In Appendix \ref{app:perf_NN} we use the same groups determined in Fig.~\ref{fig:tree}, 
but perform the classification with the NN algorithm instead of the RF algorithm.
The NN algorithm has two hidden layers with 20 and 10 nodes in the two layers respectively, tanh activation function for the hidden layers
and softmax function for the output layer.
The classification is comparable or slightly better for the NN algorithm compared to the classification with the RF algorithm.
In the next section, we use both the RF and the NN algorithm for the construction of the probabilistic catalogs.

\section{Probabilistic catalogs}
\label{sec:pcat}

\begin{table*}
\centering
\caption{Predictions for the number of associated and unassociated sources in the GMM100 catalog
calculated as a sum of class probabilities.
The class probabilities for associated (unassociated) sources are estimated as an average over testing samples (all testing / training splits).}
\label{tab:summary}
\begin{tabular}{llllllll}
\hline
 &   Node &                        Physical classes & N assoc & RF assoc & NN assoc & RF unas & NN unas \\
\hline
1 &  00000 &                     sbg, rdg, bll, css &    1514 &   1509.4 &   1529.4 &   312.4 &   267.4 \\
2 &  00100 &                              bcu, ssrq &    1495 &   1497.9 &   1490.8 &  1087.7 &  1099.2 \\
3 &  01000 &                  fsrq, sey, nlsy1, agn &     813 &    819.1 &    806.2 &   185.9 &   163.8 \\
4 &  01100 &                          glc, nov, spp &     153 &    156.4 &    151.9 &   377.4 &   419.8 \\
5 &  01101 &                               msp, lmb &     163 &    160.3 &    160.9 &   159.0 &   165.4 \\
6 &  01110 &  hmb, psr, sfr, snr, gc, pwn, gal, bin &     229 &    224.0 &    227.8 &   168.5 &   175.5 \\
\hline
\end{tabular}
\end{table*}

As a result of the analysis in the previous section, we construct three probabilistic catalogs:
\ben
\item
based on GMM group definition with $n_{\rm min} > 100$ resulting in six groups (GMM100);
\item
based on GMM group definition with $n_{\rm min} > 15$ resulting in nine groups (GMM15);
\item
based on RF group definition with $n_{\rm min} > 100$ resulting in six groups (RF100).
\een
In all catalogs we use both RF and NN algorithms for the classification with six or nine groups.
The catalogs are available online at \zenodo.
The catalogs have the 4FGL-DR3 source names, the 10 features used for the classification and definition of groups,
4FGL-DR3 source classes for associated and identified sources and probabilities for RF and NN algorithms for six- and nine-class cases respectively.
The columns with the class probabilities have the names ``m\_RF'' or ``m\_NN'' for the RF and NN probabilities respectively,
where $m = 1,\ldots, 6$ ($m = 1,\ldots, 9$) is the group index in the six (or nine) class classification.
The probabilistic catalogs and the summary of predicted numbers of sources including definitions of the groups are respectively in files:
\ben
\item GMM100:
``4FGL-DR3\_6class\_GMM\_nmin100\_prob\_cat.csv''  and ``4FGL-DR3\_6class\_GMM\_nmin100\_summary.csv'';
\item GMM15:
``4FGL-DR3\_9class\_GMM\_nmin15\_prob\_cat.csv''  and ``4FGL-DR3\_9class\_GMM\_nmin15\_summary.csv'';
\item RF100:
``4FGL-DR3\_6class\_RF\_nmin100\_prob\_cat.csv''  and ``4FGL-DR3\_6class\_RF\_nmin100\_summary.csv''.
\een
The file ``4FGL-DR3\_6class\_GMM\_nmin100\_summary.csv'' is shown in Table~\ref{tab:summary}.
It contains the index of the group, the group node name, the list of physical classes in the groups, and the predicted number of associated and unassociated sources
calculated as the sum of class probabilities for the associated and unassociated sources respectively.
For the associated sources the class probabilities are calculated as an average over the instances when the source appears in the training samples.
We require that each associated source appears at least five time in testing samples, which results in 45 random splits into training and testing samples,
i.e., on average each associated source appears 13.5 times in testing samples.
The prediction for the unassociated sources is computed as an average over 45 predictions.

\begin{table}
\centering
\caption{Examples of class probabilities for sources with largest sums of probabilities for each group of physical classes in the GMM100 catalog.}
\label{tab:likely_sources}
\begin{tabular}{llll}
\hline
       Source\_Name &                    Physical classes & RF prob & NN prob \\
\hline
 4FGL J0259.0+0552 &                    sbg, rdg, bll, css &   0.836 &   0.918 \\
 4FGL J0852.2-7208 &                             bcu, ssrq &   0.868 &   0.844 \\
 4FGL J0427.3+0504 &                 fsrq, sey, nlsy1, agn &   0.729 &   0.836 \\
4FGL J1456.4-5923c &                         glc, nov, spp &   0.735 &   0.574 \\
 4FGL J1602.2+2305 &                              msp, lmb &   0.876 &   0.917 \\
4FGL J1553.8-5325e & hmb, psr, sfr, snr, gc, &   0.896 &   0.812 \\
 &  pwn, gal, bin & & \\
\hline
\end{tabular}
\end{table}

Table~\ref{tab:likely_sources} shows the unassociated sources with the largest sums of RF and NN probabilities 
for the six classes in the GMM100 catalog.

\section{Conclusions}
\label{sec:concl}


In the paper we developed a framework for multi-class classification of \Fermi-LAT sources.
The key part of the framework is the hierarchical definition of classes, where each class consists of one or more physical classes of gamma-ray sources.
This hierarchical definition of classes has the structure of a tree, where the root contains all physical classes, and
the physical classes are separated into two groups represented by the children nodes of the root.
Such hierarchical definition of classes enables us to have control over the size of the classes used for classification and to compare
the performance for the classifications with different numbers of classes.
The possibility to compare the performance is based on the fact that the class probabilities can be computed either
directly by the ML classification for these classes or by classifying smaller classes and then adding the probabilities 
of the children nodes to obtain the class probabilities with fewer classes.
We find that the sub-division of classes does not lead to a decrease in performance (as measured by ROC curves, precision, and recall)
for the parent classes.
We also checked that the probabilities are well calibrated using reliability diagrams both for direct classification and for addition of probabilities of the children nodes.
Consequently, it is advantageous to make the classification with the smaller classes, because the probabilities of the larger classes can
be obtained by summing the probabilities of the children nodes, while the classification with more classes gives additional information about the 
likely nature of the sources.

We used two different methods (based on GMM and RF) to determine the hierarchical structure of the classes.
The classification itself was performed with RF and NN algorithms.
The classification performance was found to be similar in GMM and RF based definition of the classes and 
for RF and NN classification (the corresponding calculations are presented in appendices).
Based on the analysis in this paper, we constructed three catalogs with probabilistic classification of sources:
(1) based on GMM class definition with minimal number of sources in a group $n_{\rm min} > 100$,
(2) based on GMM class definition with minimal number of sources in a group $n_{\rm min} > 15$,
and (3) based on RF class definition with minimal number of sources in a group $n_{\rm min} > 100$.
For all catalogs we use both RF and NN algorithms for the classification of sources.
We report the probabilistic classification both for unassociated and for associated sources.
For the associated sources the probabilities are calculated as an average over instances when the sources are in the testing samples,
while for the unassociated sources the probabilities are averaged over all training-testing splits.
The catalogs are available online at \zenodo.

The probabilistic multi-class classification of \Fermi-LAT sources developed in the paper
opens up the possibility to perform population studies of different classes of the gamma-ray sources including unassociated sources,
will be useful to narrow down the possible classes of unassociated sources for follow-up observations,
and will help to understand the nature of unassociated sources, including the nature of the soft Galactic unassociated sources \citep{2022ApJS..260...53A}.

\section*{Acknowledgements}

The authors would like to thank Fabio Acero, Jean Ballet, Stefan Funk, Benoit Lott, Alison Mitchell, Ricardo Rando for useful discussions and comments.
We would like to acknowledge the use of the following software:
Astropy \citep[\url{http://www.astropy.org},][]{2013A&A...558A..33A}, 
Matplotlib \citep[\url{https://matplotlib.org/},][]{Hunter:2007}, 
pandas \citep[\url{https://pandas.pydata.org/}][]{mckinney-proc-scipy-2010},
and scikit-learn
\citep[\url{https://scikit-learn.org/stable/},][]{scikit-learn}.


\section*{Data Availability}

The results of this work are based on the publicly available \Fermi-LAT 4FGL-DR3 catalog \url{https://fermi.gsfc.nasa.gov/ssc/data/access/lat/12yr_catalog/}
\citep{2022ApJS..260...53A}.
The results of this work are available online at \zenodo.
 



\bibliographystyle{mnras}
\bibliography{Fermi_multi_bibl} 

\begin{thebibliography}{}
\makeatletter
\relax
\def\mn@urlcharsother{\let\do\@makeother \do\$\do\&\do\#\do\^\do\_\do\%\do\~}
\def\mn@doi{\begingroup\mn@urlcharsother \@ifnextchar [ {\mn@doi@}
  {\mn@doi@[]}}
\def\mn@doi@[#1]#2{\def\@tempa{#1}\ifx\@tempa\@empty \href
  {http://dx.doi.org/#2} {doi:#2}\else \href {http://dx.doi.org/#2} {#1}\fi
  \endgroup}
\def\mn@eprint#1#2{\mn@eprint@#1:#2::\@nil}
\def\mn@eprint@arXiv#1{\href {http://arxiv.org/abs/#1} {{\tt arXiv:#1}}}
\def\mn@eprint@dblp#1{\href {http://dblp.uni-trier.de/rec/bibtex/#1.xml}
  {dblp:#1}}
\def\mn@eprint@#1:#2:#3:#4\@nil{\def\@tempa {#1}\def\@tempb {#2}\def\@tempc
  {#3}\ifx \@tempc \@empty \let \@tempc \@tempb \let \@tempb \@tempa \fi \ifx
  \@tempb \@empty \def\@tempb {arXiv}\fi \@ifundefined
  {mn@eprint@\@tempb}{\@tempb:\@tempc}{\expandafter \expandafter \csname
  mn@eprint@\@tempb\endcsname \expandafter{\@tempc}}}

\bibitem[\protect\citeauthoryear{{Abdo}, {Ackermann}, {Ajello}  et~al.}{{Abdo}
  et~al.}{2010}]{2010ApJS..188..405A}
{Abdo} A.~A.,  {Ackermann} M.,  {Ajello} M.,   et~al., 2010, \mn@doi [ApJS]
  {10.1088/0067-0049/188/2/405}, \href
  {https://ui.adsabs.harvard.edu/abs/2010ApJS..188..405A} {188, 405}

\bibitem[\protect\citeauthoryear{{Abdollahi}, {Acero}, {Ackermann}
  et~al.}{{Abdollahi} et~al.}{2020}]{2020ApJS..247...33A}
{Abdollahi} S.,  {Acero} F.,  {Ackermann} M.,   et~al., 2020, \mn@doi [ApJS]
  {10.3847/1538-4365/ab6bcb}, \href
  {https://ui.adsabs.harvard.edu/abs/2020ApJS..247...33A} {247, 33}

\bibitem[\protect\citeauthoryear{{Abdollahi} et~al.,}{{Abdollahi}
  et~al.}{2022}]{2022ApJS..260...53A}
{Abdollahi} S.,  et~al., 2022, \mn@doi [\apjs] {10.3847/1538-4365/ac6751},
  \href {https://ui.adsabs.harvard.edu/abs/2022ApJS..260...53A} {260, 53}

\bibitem[\protect\citeauthoryear{{Acero}, {Ackermann}, {Ajello}
  et~al.}{{Acero} et~al.}{2015}]{2015ApJS..218...23A}
{Acero} F.,  {Ackermann} M.,  {Ajello} M.,   et~al., 2015, \mn@doi [ApJS]
  {10.1088/0067-0049/218/2/23}, \href
  {http://adsabs.harvard.edu/abs/2015ApJS..218...23A} {218, 23}

\bibitem[\protect\citeauthoryear{{Ackermann}, {Ajello}, {Allafort}
  et~al.}{{Ackermann} et~al.}{2012}]{2012ApJ...753...83A}
{Ackermann} M.,  {Ajello} M.,  {Allafort} A.,   et~al., 2012, \mn@doi [ApJ]
  {10.1088/0004-637X/753/1/83}, \href
  {https://ui.adsabs.harvard.edu/abs/2012ApJ...753...83A} {753, 83}

\bibitem[\protect\citeauthoryear{{Bhat} \& {Malyshev}}{{Bhat} \&
  {Malyshev}}{2022}]{2022A&A...660A..87B}
{Bhat} A.,  {Malyshev} D.,  2022, \mn@doi [A\&A] {10.1051/0004-6361/202140766},
  \href {https://ui.adsabs.harvard.edu/abs/2022A&A...660A..87B} {660, A87}

\bibitem[\protect\citeauthoryear{{Coronado-Bl{\'a}zquez}}{{Coronado-Bl{\'a}zquez}}{2022}]{2022MNRAS.515.1807C}
{Coronado-Bl{\'a}zquez} J.,  2022, \mn@doi [\mnras] {10.1093/mnras/stac1950},
  \href {https://ui.adsabs.harvard.edu/abs/2022MNRAS.515.1807C} {515, 1807}

\bibitem[\protect\citeauthoryear{{Finke}, {Kr{\"a}mer}  \& {Manconi}}{{Finke}
  et~al.}{2021}]{2021MNRAS.507.4061F}
{Finke} T.,  {Kr{\"a}mer} M.,   {Manconi} S.,  2021, \mn@doi [MNRAS]
  {10.1093/mnras/stab2389}, \href
  {https://ui.adsabs.harvard.edu/abs/2021MNRAS.507.4061F} {507, 4061}

\bibitem[\protect\citeauthoryear{Hunter}{Hunter}{2007}]{Hunter:2007}
Hunter J.~D.,  2007, \mn@doi [Computing In Science \& Engineering]
  {10.1109/MCSE.2007.55}, 9, 90

\bibitem[\protect\citeauthoryear{{Lefaucheur} \& {Pita}}{{Lefaucheur} \&
  {Pita}}{2017}]{2017A&A...602A..86L}
{Lefaucheur} J.,  {Pita} S.,  2017, \mn@doi [A\&A]
  {10.1051/0004-6361/201629552}, \href
  {https://ui.adsabs.harvard.edu/abs/2017A&A...602A..86L} {602, A86}

\bibitem[\protect\citeauthoryear{{Luo}, {Leung}, {Hui}  \& {Li}}{{Luo}
  et~al.}{2020}]{2020MNRAS.492.5377L}
{Luo} S.,  {Leung} A.~P.,  {Hui} C.~Y.,   {Li} K.~L.,  2020, \mn@doi [MNRAS]
  {10.1093/mnras/staa166}, \href
  {https://ui.adsabs.harvard.edu/abs/2020MNRAS.492.5377L} {492, 5377}

\bibitem[\protect\citeauthoryear{{Malyshev} \& {Bhat}}{{Malyshev} \&
  {Bhat}}{2022}]{2022arXiv220910236M}
{Malyshev} D.,  {Bhat} A.,  2022, \href
  {https://ui.adsabs.harvard.edu/abs/2022arXiv220910236M} {}

\bibitem[\protect\citeauthoryear{{Mirabal}, {Charles}, {Ferrara}, {Gonthier},
  {Harding}, {S{\'a}nchez-Conde}  \& {Thompson}}{{Mirabal}
  et~al.}{2016}]{2016ApJ...825...69M}
{Mirabal} N.,  {Charles} E.,  {Ferrara} E.~C.,  {Gonthier} P.~L.,  {Harding}
  A.~K.,  {S{\'a}nchez-Conde} M.~A.,   {Thompson} D.~J.,  2016, \mn@doi [ApJ]
  {10.3847/0004-637X/825/1/69}, \href
  {https://ui.adsabs.harvard.edu/abs/2016ApJ...825...69M} {825, 69}

\bibitem[\protect\citeauthoryear{{Nolan}, {Abdo}, {Ackermann}  et~al.}{{Nolan}
  et~al.}{2012}]{2012ApJS..199...31N}
{Nolan} P.~L.,  {Abdo} A.~A.,  {Ackermann} M.,   et~al., 2012, \mn@doi [ApJS]
  {10.1088/0067-0049/199/2/31}, \href
  {http://adsabs.harvard.edu/abs/2012ApJS..199...31N} {199, 31}

\bibitem[\protect\citeauthoryear{Pedregosa et~al.,}{Pedregosa
  et~al.}{2011}]{scikit-learn}
Pedregosa F.,  et~al., 2011, Journal of Machine Learning Research, 12, 2825

\bibitem[\protect\citeauthoryear{{Robitaille}, {Tollerud}, {Greenfield}
  et~al.}{{Robitaille} et~al.}{2013}]{2013A&A...558A..33A}
{Robitaille} T.~P.,  {Tollerud} E.~J.,  {Greenfield} P.,   et~al., 2013,
  \mn@doi [A\&A] {10.1051/0004-6361/201322068}, \href
  {http://adsabs.harvard.edu/abs/2013A%26A...558A..33A} {558, A33}

\bibitem[\protect\citeauthoryear{{Saz Parkinson}, {Xu}, {Yu}, {Salvetti},
  {Marelli}  \& {Falcone}}{{Saz Parkinson} et~al.}{2016}]{2016ApJ...820....8S}
{Saz Parkinson} P.~M.,  {Xu} H.,  {Yu} P.~L.~H.,  {Salvetti} D.,  {Marelli} M.,
    {Falcone} A.~D.,  2016, \mn@doi [ApJ] {10.3847/0004-637X/820/1/8}, \href
  {https://ui.adsabs.harvard.edu/abs/2016ApJ...820....8S} {820, 8}

\bibitem[\protect\citeauthoryear{{Sokolova} \& {Rubtsov}}{{Sokolova} \&
  {Rubtsov}}{2016}]{2016ApJ...833..271S}
{Sokolova} E.~V.,  {Rubtsov} G.~I.,  2016, \mn@doi [\apj]
  {10.3847/1538-4357/833/2/271}, \href
  {https://ui.adsabs.harvard.edu/abs/2016ApJ...833..271S} {833, 271}

\bibitem[\protect\citeauthoryear{{W}es {M}c{K}inney}{{W}es
  {M}c{K}inney}{2010}]{mckinney-proc-scipy-2010}
{W}es {M}c{K}inney 2010, in {S}t\'efan van~der {W}alt {J}arrod {M}illman eds,
  {P}roceedings of the 9th {P}ython in {S}cience {C}onference. pp 56 -- 61,
  \mn@doi{10.25080/Majora-92bf1922-00a}

\bibitem[\protect\citeauthoryear{{Zhu}, {Kang}  \& {Zheng}}{{Zhu}
  et~al.}{2021}]{2021RAA....21...15Z}
{Zhu} K.-R.,  {Kang} S.-J.,   {Zheng} Y.-G.,  2021, \mn@doi [Research in
  Astronomy and Astrophysics] {10.1088/1674-4527/21/1/15}, \href
  {https://ui.adsabs.harvard.edu/abs/2021RAA....21...15Z} {21, 015}

\makeatother
\end{thebibliography}



\newpage
\appendix

\section{Groups with smaller minimal size}
\label{app:nmin15}

\begin{figure}
\includegraphics[width=\columnwidth]{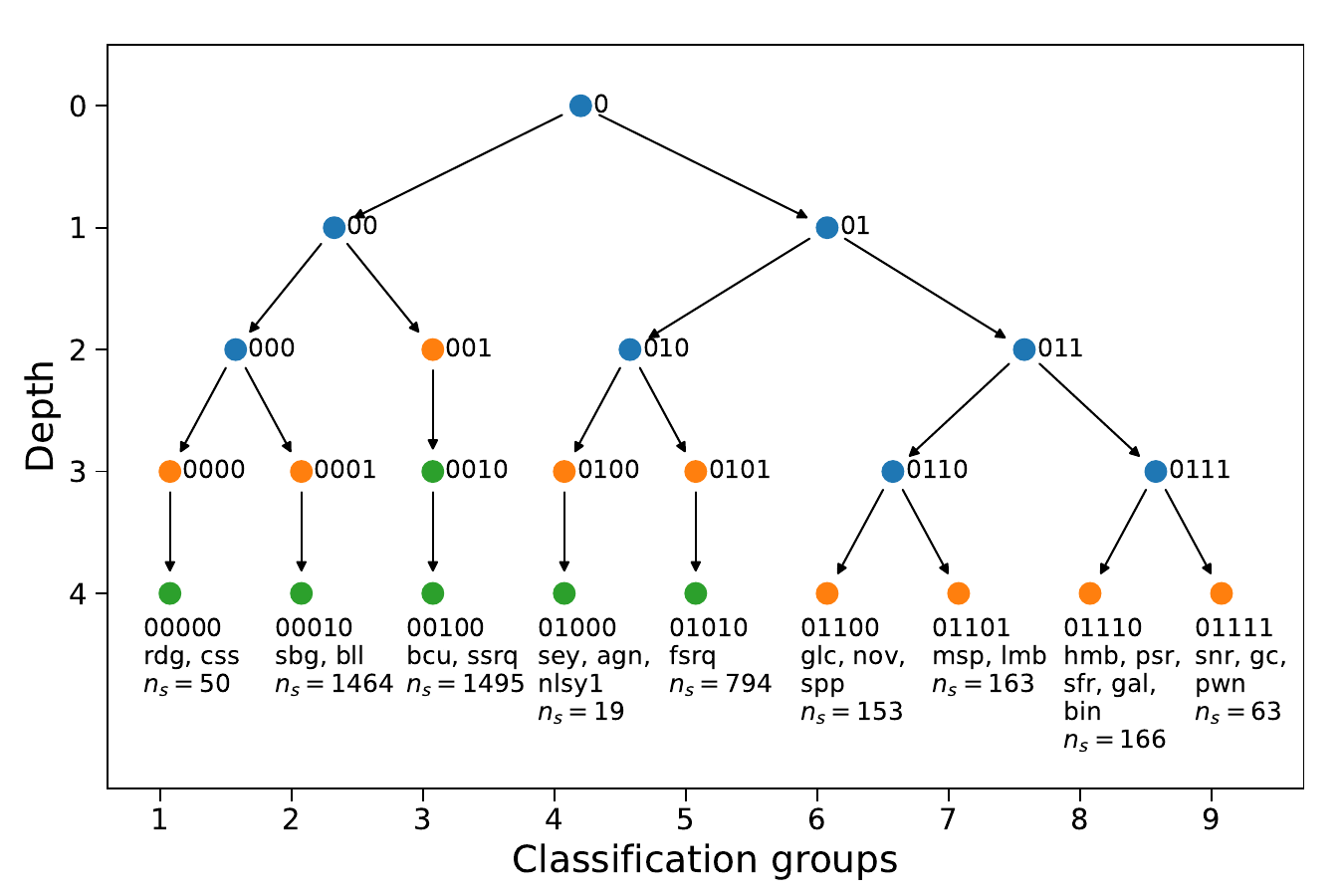}
\caption{Determination of the groups of classes using the GMM algorithm with the minimal number of 
sources in a group $n_{\rm min} > 15$.
The algorithm is described in Section \ref{sec:class_def}.
The nodes are defined similarly to Figure \ref{fig:tree}.
}
\label{fig:tree_nmin15}
\end{figure}

\begin{figure*}
\includegraphics[width=2\columnwidth]{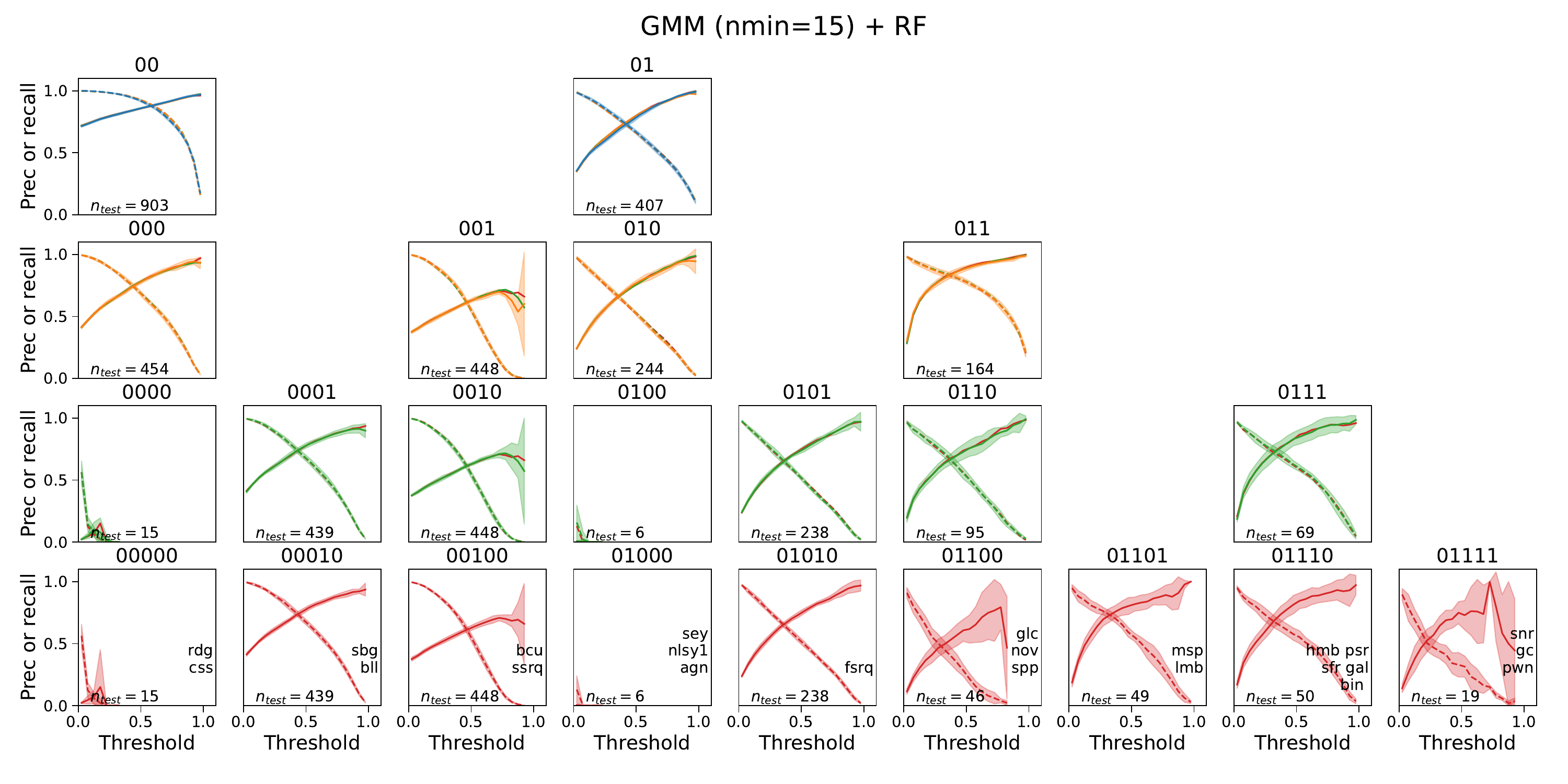}
\caption{Precision and recall for all groups in Figure \ref{fig:tree_nmin15}. The lines are defined similarly to Figure \ref{fig:prec_recall_all}.}
\label{fig:pr_all_nmin15}
\end{figure*}

In this appendix we perform classification including groups of smaller minimal size $n_{\rm min} > 15$
compared to the condition $n_{\rm min} > 100$ used in Sections \ref{sec:method} and \ref{sec:classif}.
The separation of classes using the GMM method with $n_{\rm min} > 15$ and maximal depth four is shown in Figure \ref{fig:tree_nmin15}.
Similarly to Section \ref{sec:classif}, we use the RF algorithm for the classification.
The precision and recall for all groups in Figure \ref{fig:tree_nmin15} are shown in Figure \ref{fig:pr_all_nmin15}.
We see that, as in the case of $n_{\rm min} > 100$ condition, the precision and recall are similar for direct classification and for
classification with more classes and summation of probabilities of children nodes.
Most of groups have reasonable precision and recall, 
apart from the two smallest groups 00000 and 00100 with 50 and 19 associated sources respectively.
Interestingly, the group 01111 has 63 associated sources, which is just 13 sources more than the group 00000,
but shows much nicer precision and recall, which means that even relatively small groups can be recovered,
if they are well separated from the larger groups in the feature space.

\begin{figure*}
\includegraphics[width=\cmpsize\columnwidth]{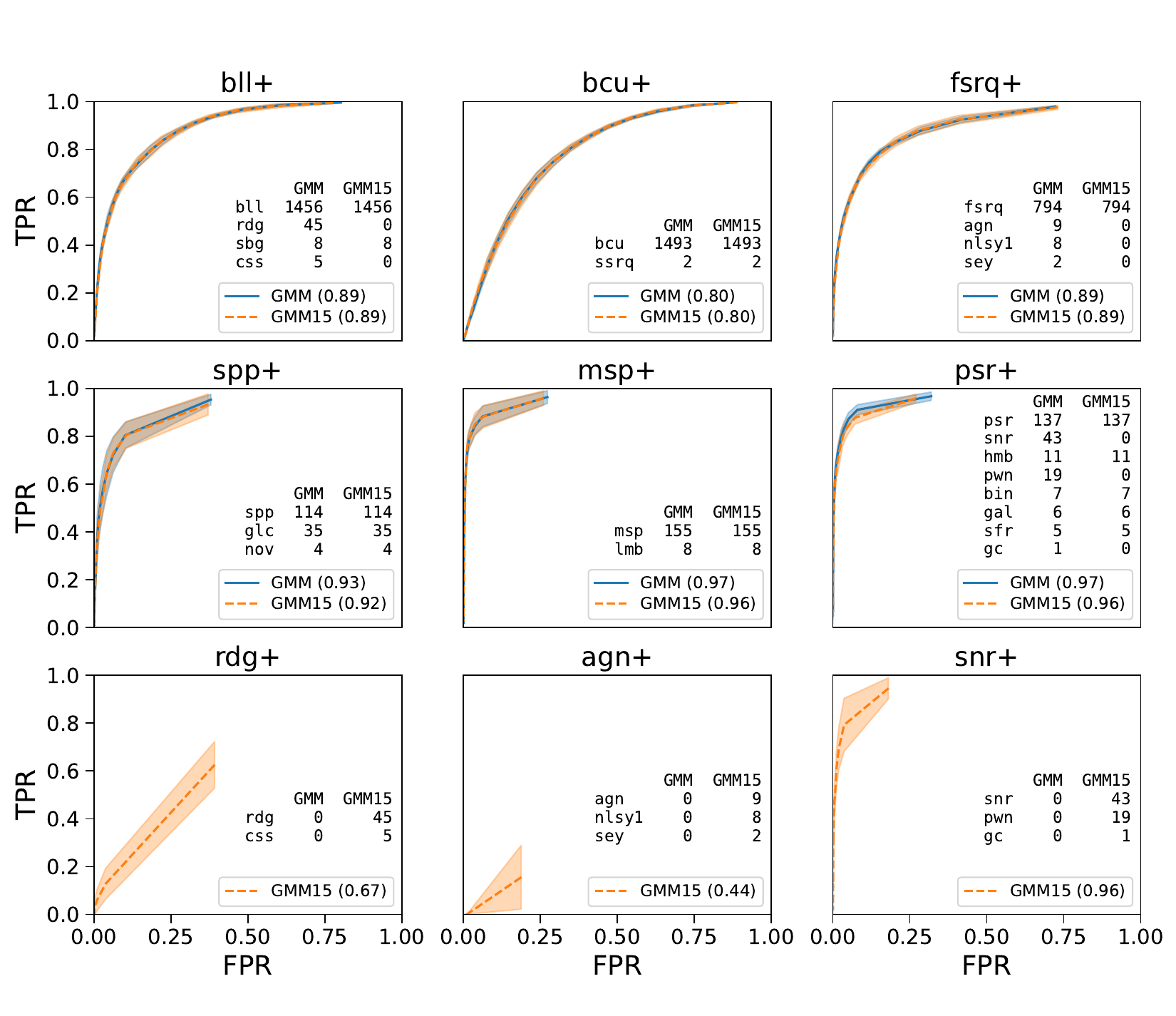}
\caption{Comparison of ROC curves for the groups defined with the GMM with minimal number of sources in a group $n_{\rm min} > 100$ (the GMM label)
and the groups defined with GMM and $n_{\rm min} > 15$ (the GMM15 label).
The panels in the first two rows correspond to groups with a common large physical class, e.g., bll, bcu, fsrq, spp, msp, and psr respectively.
The last row shows the groups with largest physical classes rdg, agn, and snr, which are independent in the nine-class classification with $n_{\rm min} > 15$, 
but belong to larger groups in the six-class classification with  $n_{\rm min} > 100$.
The tables on the panels show the physical classes, which belong to the corresponding groups and the numbers of sources in these classes.
The numbers in parentheses show the areas under the curves.
Shaded areas show the statistical uncertainty estimated by 10 random splits into training and testing data.
}
\label{fig:ROC_compare_nmin15}
\end{figure*}

We compare the ROC curves for the nine-class classification in the $n_{\rm min} > 15$ case and the six-class classification
in the $n_{\rm min} > 100$ case in Figure \ref{fig:ROC_compare_nmin15}.
Since the groups have different combinations of physical classes, direct comparison of performances is not possible.
Nevertheless, assuming that the classification performance is driven by the largest physical class in the group,
we compare the ROC curves for the classes which contain the same largest class.
In the $n_{\rm min} > 100$ case, the six largest classes are bll, bcu, fsrq, spp, msp, and psr.
These are also the largest classes in six out of nine groups in the nine-class classification.
The corresponding comparison of the ROC curves is shown in the first two rows in Figure \ref{fig:ROC_compare_nmin15}.
In the remaining three groups in the nine-class classification, the largest classes are rdg, agn, and snr 
(shown in the last row in Figure \ref{fig:ROC_compare_nmin15}).
The tables in the panels show the number of sources in each of the groups,
where GMM labels the baseline six-class classification with $n_{\rm min} > 100$
while GMM15 labels the nine-class classification with $n_{\rm min} > 15$.
Shaded blue (orange) areas show statistical uncertainties estimated from 10 random training / testing
splits in the GMM (GMM15) cases.
Numbers in parentheses in the labels show the areas under curve.
The performance is slightly worse for the GMM15 case for the psr+ class, 
but it has an additional snr+ class with a very good area under the curve of 0.97.

\begin{figure*}
\includegraphics[width=\cmpsize\columnwidth]{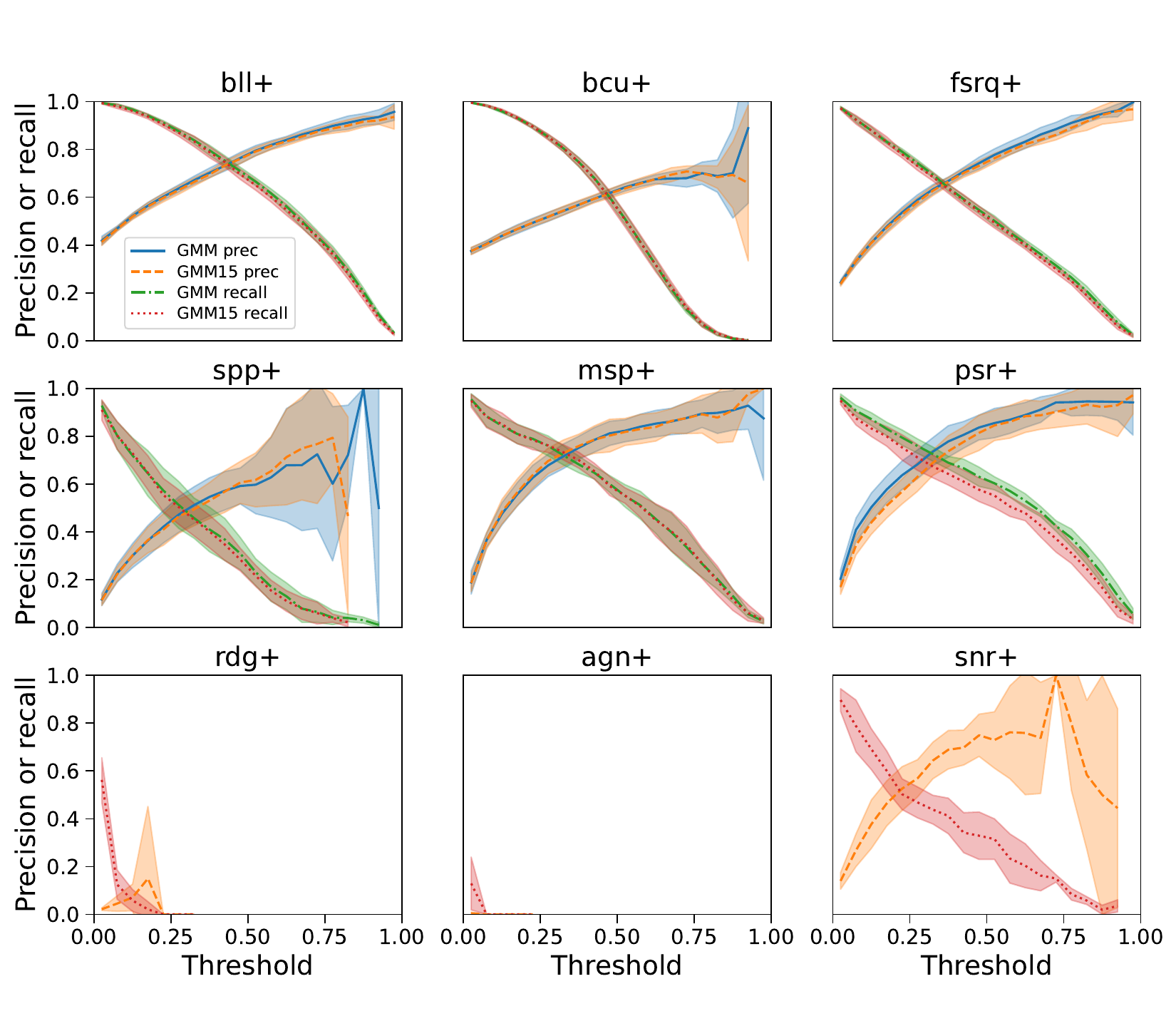}
\caption{Comparison of precision and recall for the groups defined with the GMM with minimal number of sources in a group $n_{\rm min} > 100$ (the GMM label)
and the groups defined with GMM and $n_{\rm min} > 15$ (the GMM15 label).
The groups have the same definition as in Figure \ref{fig:ROC_compare_nmin15}.}
\label{fig:pr_compare_nmin15}
\end{figure*}

The precision and recall for the nine-class and the six-class classifications are compared in Figure \ref{fig:pr_compare_nmin15}
similarly to the comparison of the ROC curves in Figure \ref{fig:ROC_compare_nmin15}.
We find that the precision and recall have similar values for the groups dominated by the common largest physical class shown in the 
first two rows in Figure \ref{fig:pr_compare_nmin15} with a slightly worse performance for the psr+ case, 
but with a new group snr+ in the nine-class case, which also has reasonable precision and recall.
The precision and recall for the two smallest groups rdg+ and agn+ are close to zero (in the rdg+ case, one can recover some rdg or css sources 
with probabilities less than about 0.1).

We have also checked that the ROC curves for all groups in Figure \ref{fig:tree_nmin15} 
show a similar behaviour as in Figure \ref{fig:ROC_all}
apart from the two smallest groups, for which the ROC curves
are consistent with random guess (as shown in Figure \ref{fig:ROC_compare_nmin15})
while precision and recall are close to zero (as in Figure \ref{fig:pr_compare_nmin15}).
We have also used reliability diagrams to check that the probabilities are well calibrated for all groups in Figure \ref{fig:tree_nmin15}, 
similarly to Figure \ref{fig:reliability}.
The ROC curves and the reliability diagrams are not shown for the sake of brevity.
Overall, we find that the nine-class classification provides a similar performance to the six-class classification 
for groups with the common largest physical class with one additional
class (snr+), which has reasonable ROC curves, precision and recall.
As a result, we also create a catalog with the nine-class classification with $n_{\rm min} > 15$ 
(the GMM15 catalog in Section \ref{sec:pcat}), in addition to the six-class catalog with $n_{\rm min} > 100$
(labeled as GMM100 catalog in Section \ref{sec:pcat}).

\section{Definition of groups with random forest}
\label{app:RFclasses}

\begin{figure}
\includegraphics[width=\columnwidth]{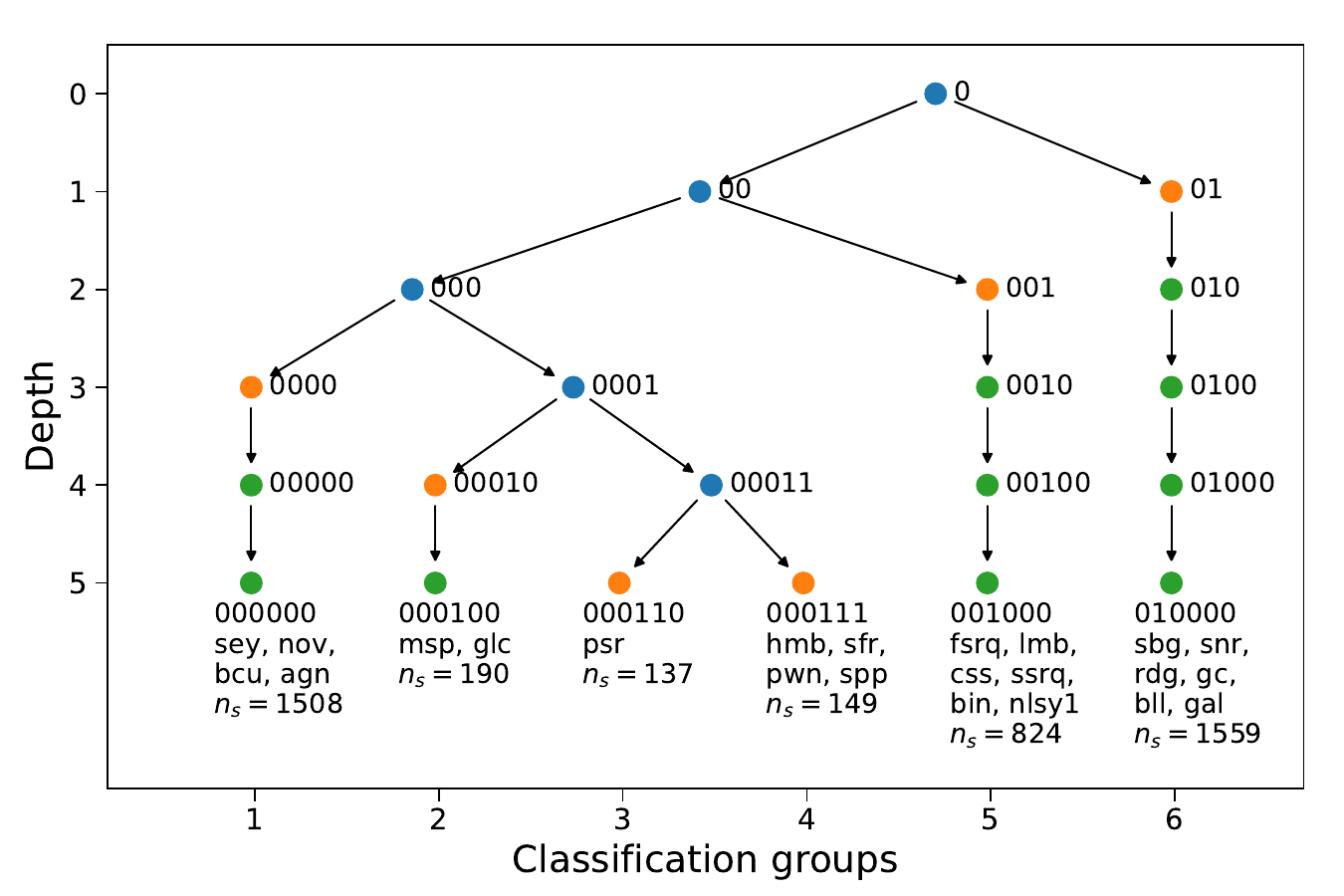}
\caption{
Determination of the groups of classes using the RF algorithm with minimal number of 
sources in a group $n_{\rm min} > 100$.
The nodes are defined similarly to Figure \ref{fig:tree}.
}
\label{fig:tree_RF}
\end{figure}

In this appendix we define the groups of physical classes using the RF algorithm instead of the GMM.
In order to determine the hierarchical structure of the groups, we first select two largest classes among all associated sources, bcu and bll,
and train the RF to classify all associated sources in these two classes, denoted as class 0 and 1 respectively.
Then we proceed similarly to the GMM case.
We calculate the class 0 and class 1 probabilities for all associated sources and attribute a physical class to class 1 if the average class-1
probability for the sources in this physical class is larger than the average class-1 probability for all associated sources
(otherwise the physical class is attributed to class 0).
We repeat this procedure iteratively for the sub-groups by selecting two largest classes in each of the subgroups to train the RF algorithm
for the separation in this node.
We use the condition on the minimal number of sources in a group $n_{\rm min} > 100$.
The resulting separation of the physical classes into groups is presented in Figure \ref{fig:tree_RF}.

\begin{figure*}
\includegraphics[width=\allsize\columnwidth]{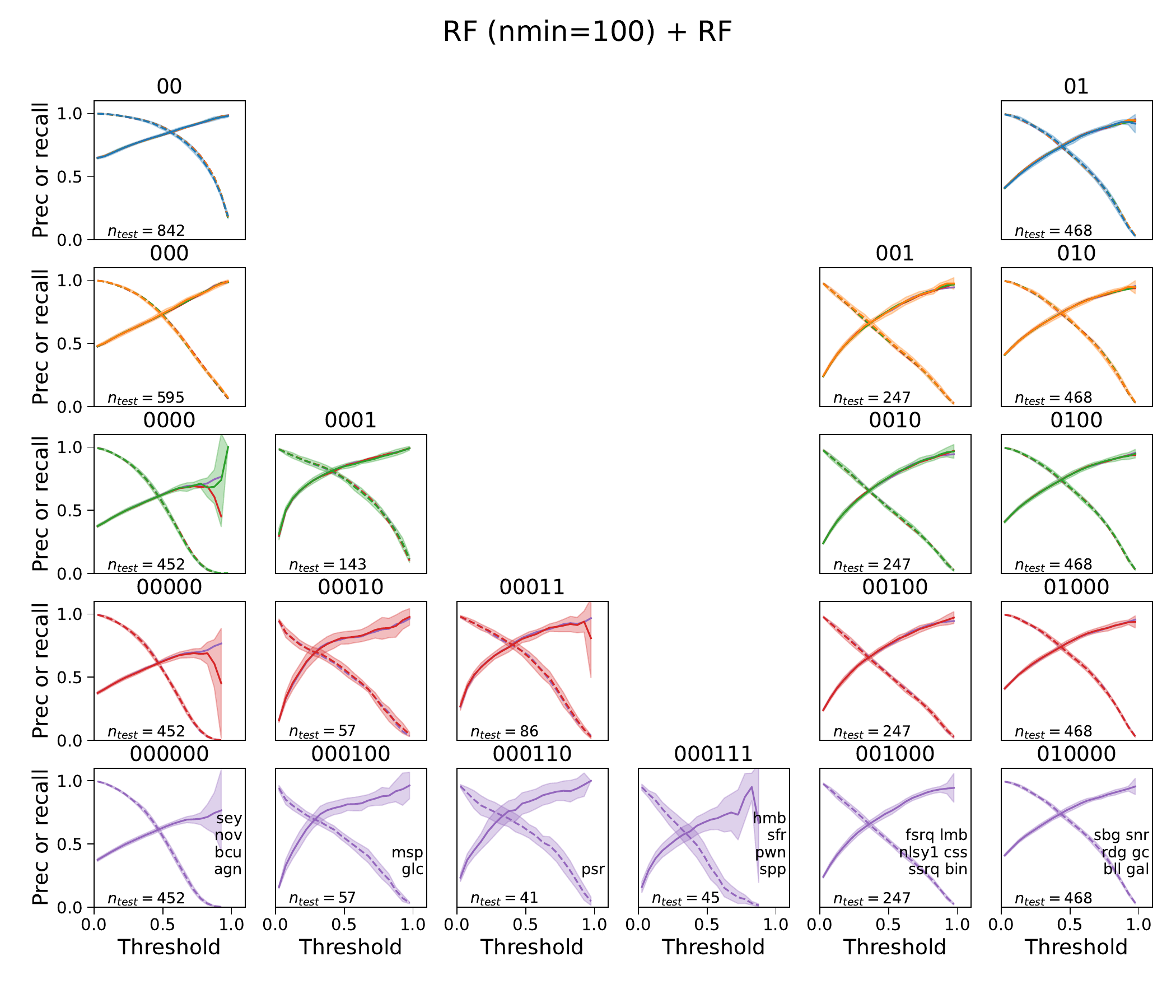}
\caption{Precision and recall for all groups in Figure \ref{fig:tree_RF}. The lines are defined similarly to Figure \ref{fig:prec_recall_all}.}
\label{fig:pr_all_RF}
\end{figure*}

We use the same RF algorithm for the classification as in Section \ref{sec:classif}.
We show the precision and recall in Figure \ref{fig:pr_all_RF} for all groups in Figure \ref{fig:tree_RF}.
Similarly to the GMM definitions of the groups in Section \ref{sec:class_def} and in Appendix \ref{app:nmin15},
the precision and recall are similar for direct classification and for classification with more groups and with summation of the class probabilities 
of the children nodes.

\begin{figure*}
\includegraphics[width=\cmpsize\columnwidth]{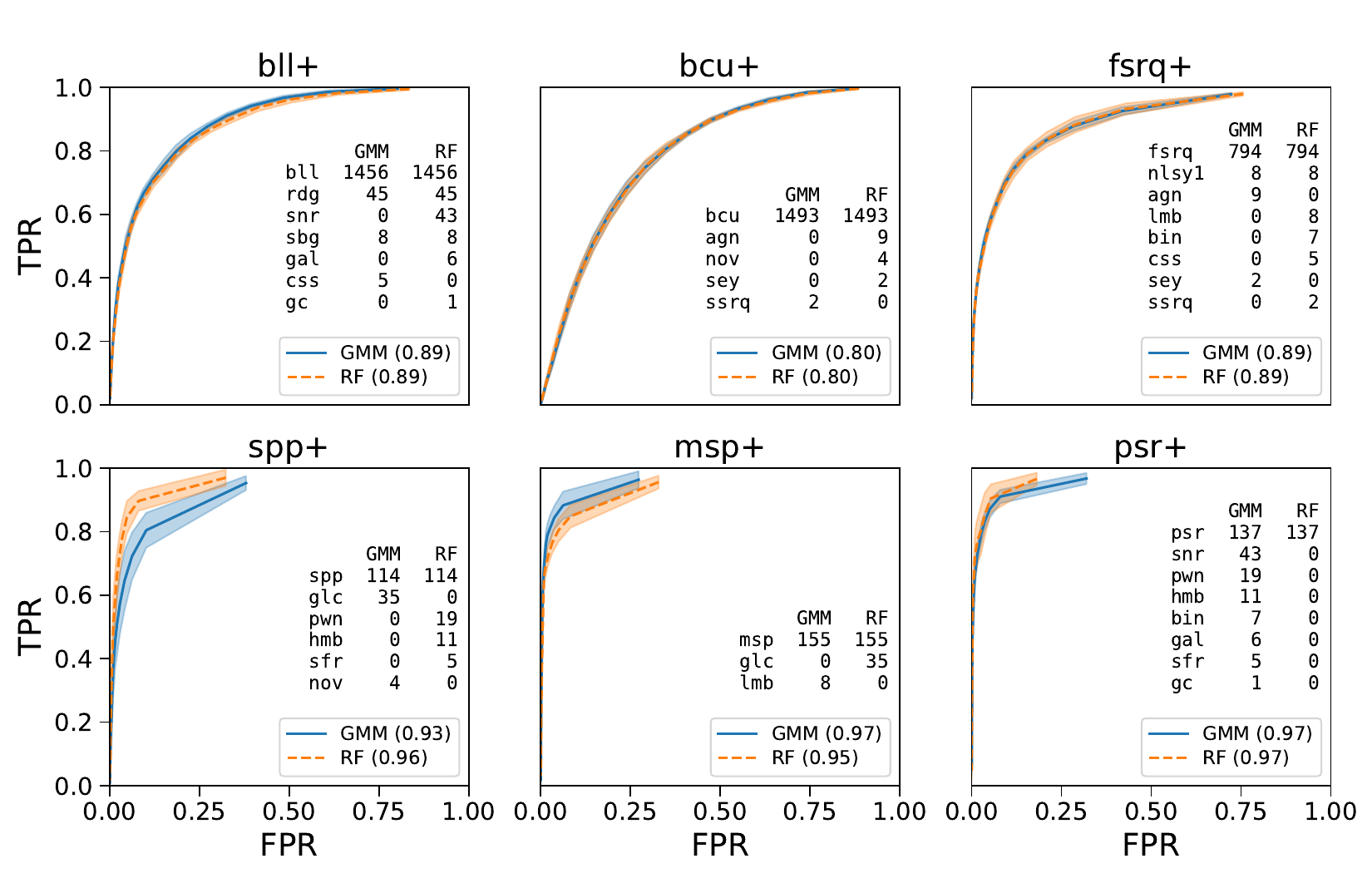}
\caption{Comparison of ROC curves for the groups defined with the GMM (the GMM labels)
and the groups defined with the RF (the RF labels) algorithms. In both cases the minimal number of sources in a group $n_{\rm min} > 100$.
The panels show the comparison of groups with a common large physical class, e.g., bll, bcu, fsrq, spp, msp, and psr respectively.
The lines and the tables in the panels are defined similar to Figure \ref{fig:ROC_compare_nmin15}.
}
\label{fig:ROC_compare_RF}
\end{figure*}

In Figure \ref{fig:ROC_compare_RF}, we compare the ROC curves for the GMM definition of the groups (GMM labels) 
 with the RF definition of the groups (RF labels).
In both cases, we use $n_{\rm min} > 100$ condition for the minimal number of sources in a group, which results in six classes at the largest depth with the 
same maximal physical classes, but with different distributions of smaller classes among the groups.
Also in both cases, the RF algorithm is used for the classification.
In Figure \ref{fig:ROC_compare_RF}, we compare the ROC curves for the groups which have the same maximal physical class.
We see that the groups determined with the RF algorithm have a similar performance as the groups determined with the GMM method,
apart from the spp+ group, where the RF-based group has a better ROC curve compared to the GMM based spp+ group.

\begin{figure*}
\includegraphics[width=\cmpsize\columnwidth]{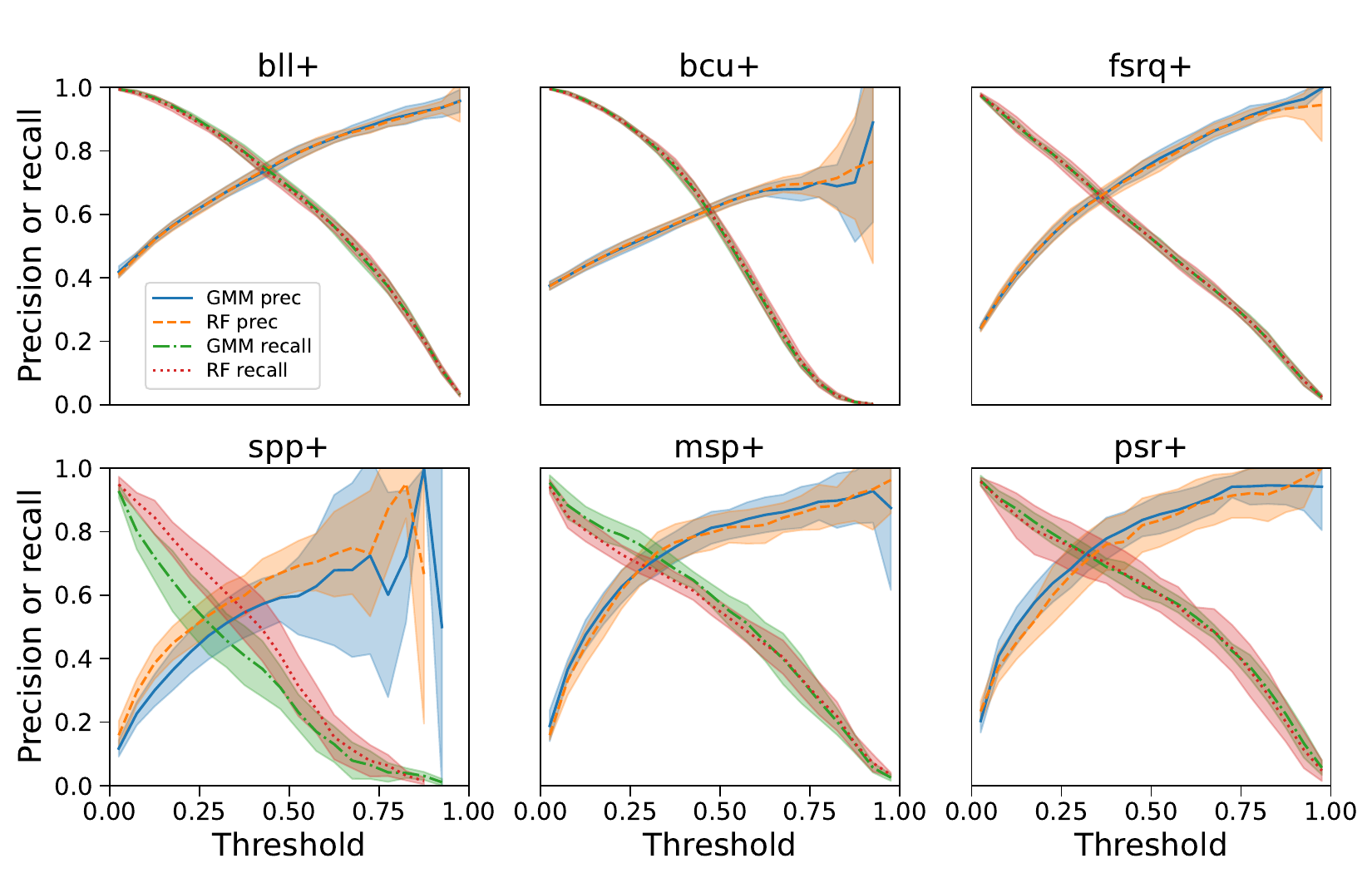}
\caption{
Comparison of precision and recall for the groups defined with the GMM (the GMM labels)
and the groups defined with the RF (the RF labels) algorithms. 
The groups have the same definition as in Figure \ref{fig:ROC_compare_RF}.}
\label{fig:pr_compare_RF}
\end{figure*}

In Figure \ref{fig:pr_compare_RF}, we compare the precision and recall for the GMM definition of the groups (GMM labels) 
 with the RF definition of the groups (RF labels).
The groups are the same as in Figure \ref{fig:ROC_compare_RF}.
The lines are defined similar to Figure \ref{fig:pr_compare_nmin15}.
Similarly to the ROC curves,
the precision and recall are comparable for the groups determined with the RF algorithm and for the groups determined with the GMM method,
apart from the spp+ group, where the RF-based group has better precision and recall compared to the GMM based spp+ group.
Since the classification based on the RF determination of the groups has a similar or better performance compared to the groups determined with the GMM method,
we also create a catalog based on the RF determination of the groups (the RF100 catalog in Section \ref{sec:pcat}).

\section{Classification with neural networks}
\label{app:perf_NN}

In this appendix we compare the performance of the classification using RF algorithm (Section \ref{sec:classif}) and the NN algorithm.
We use the same groups as in Section \ref{sec:class_def}, Figure \ref{fig:tree}.
The NN has the same 10 input features as the RF algorithm in Section \ref{sec:classif} and
two hidden layers with 20 and 10 nodes respectively.
The number of output nodes is equal to the number of classes (two, four, five, or six).
The activation function after the hidden layers is tanh, the activation function for the output layer is softmax.
We use the ``MLPClassifier'' implementation of NN in scikit-learn v1.0.2 \citep{scikit-learn}.
The objective function is log-loss (cross-entropy).

\begin{figure*}
\includegraphics[width=\allsize\columnwidth]{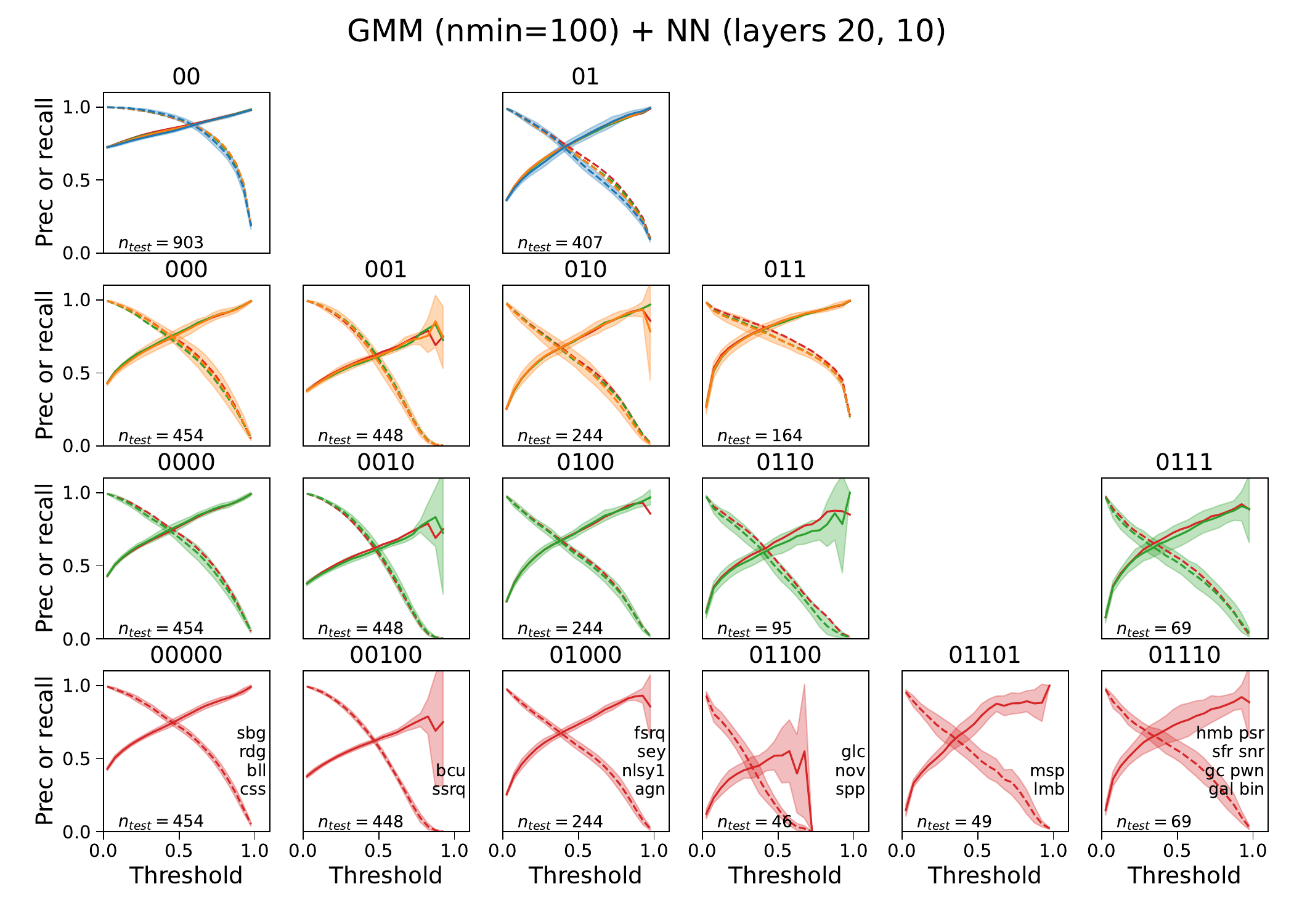}
\caption{Precision and recall for all groups in Figure \ref{fig:tree} calculated with the NN method. 
The lines are defined similarly to Figure \ref{fig:prec_recall_all}.}
\label{fig:pr_all_NN}
\end{figure*}

The precision and recall for all groups in Figure \ref{fig:tree} using the NN classification is shown in Figure \ref{fig:pr_all_NN}.
Similarly to the RF classification considered earlier,
the precision and recall are similar for direct classification and for classification with more groups and with summation of the class probabilities 
of the children nodes.
Although the deviations are more visible now for some of the nodes, e.g., 01, 0110, and 0111, where classification in subgroups and summation of probabilities
gives slightly better performance compared to direct classification,
which reinforces the conclusion that it is advantageous to perform a classification with smaller groups.

\begin{figure*}
\includegraphics[width=\cmpsize\columnwidth]{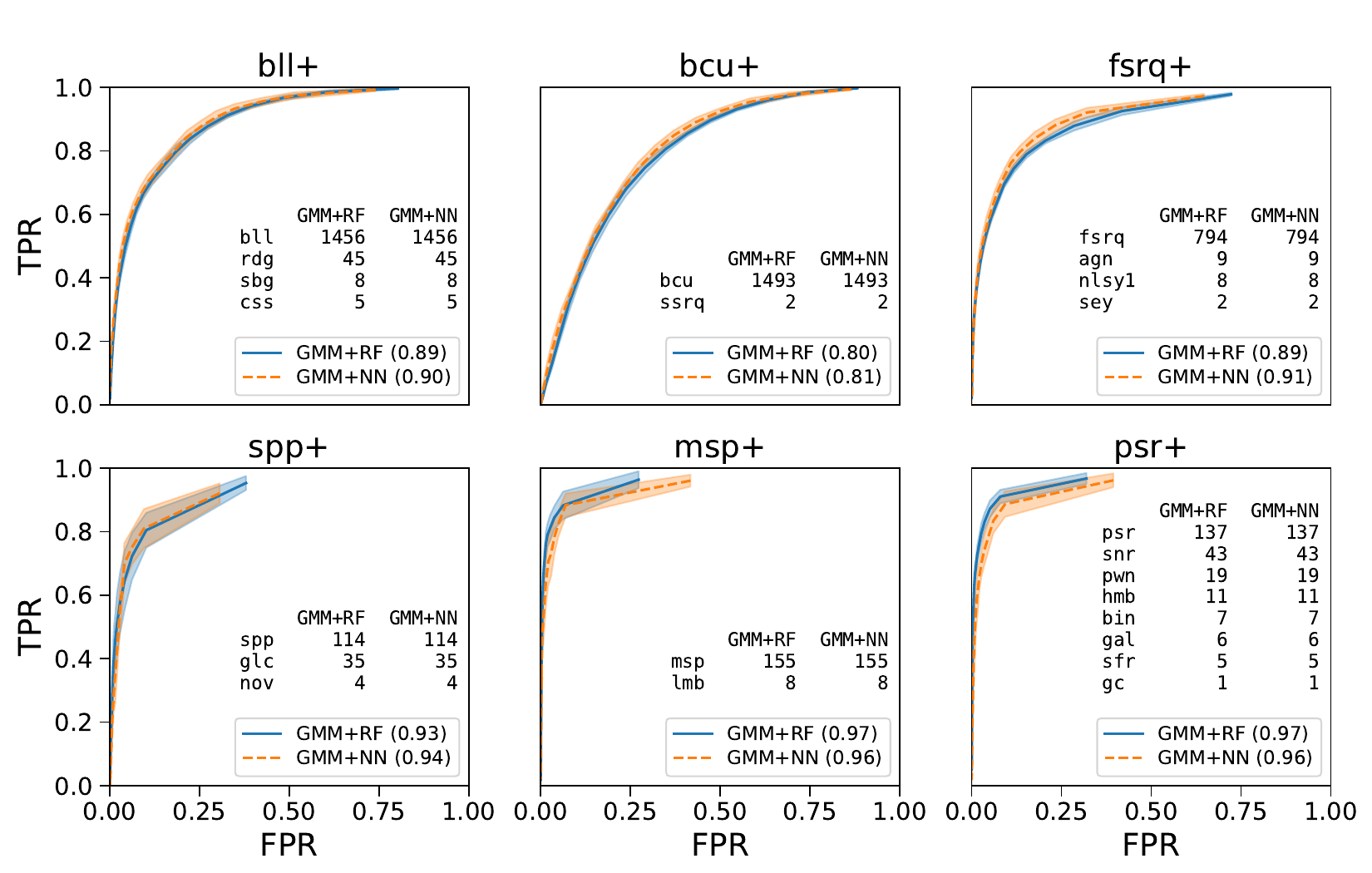}
\caption{Comparison of ROC curves for the classification of the groups at depth four in Figure \ref{fig:tree} with the RF method (GMM+RF labels) 
and with the NN method (GMM+NN labels).
In both cases the classes are the same and the only difference is in the classification method.
The lines and the tables in the panels are defined similar to Figure \ref{fig:ROC_compare_nmin15}.
}
\label{fig:ROC_compare_NN}
\end{figure*}

\begin{figure*}
\includegraphics[width=\cmpsize\columnwidth]{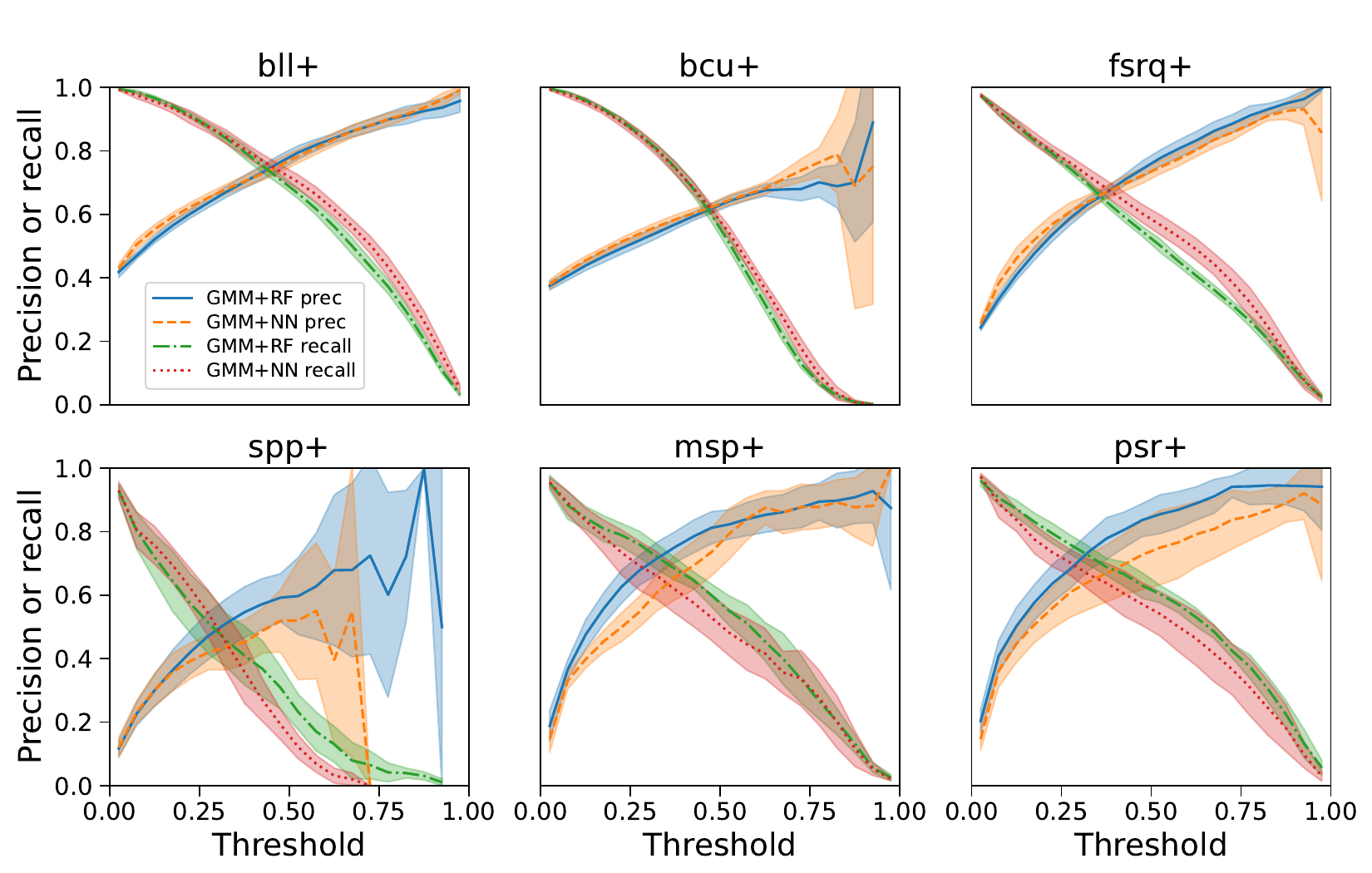}
\caption{Comparison of precision and recall for the classification of the groups at depth four in Figure \ref{fig:tree} with the RF method (GMM+RF labels) 
and with the NN method (GMM+NN labels).
The groups have the same definition as in Figure \ref{fig:ROC_compare_NN}.}
\label{fig:pr_compare_NN}
\end{figure*}

The ROC curves for the RF and the NN classifications are compared in Figure \ref{fig:ROC_compare_NN}.
For most of the groups the area under the curve is slightly better for the NN classification compared to the RF one,
with the exception of the psr+ group, where the RF classification has a slightly better performance.
The precision and recall for the RF and the NN classifications are compared in Figure \ref{fig:pr_compare_NN}.
Also in this case the precision and recall for the RF and NN classifications are similar, 
where for different intervals in the probability threshold either RF or NN methods have a slightly better performance.
Overall, RF and NN classification provide comparable ROC curves, precision, and recall for the same groups.
As a result, we report both the RF and the NN probabilities in the probabilistic catalogs constructed in Section \ref{sec:pcat}.

\bsp	
\label{lastpage}
\end{document}